\documentclass[useAMS,usenatbib]{mn2e}
\usepackage{times}
\usepackage{natbib}
\usepackage{aas_macros}
\usepackage{amsmath}
\usepackage[pdftex]{graphicx}

\title[]{Initial Conditions for Idealised Clusters Mergers, simulating ''El Gordo''}
\author[J. M. F. Donnert]{
J. M. F. Donnert$^{1}$,\thanks{donnert@ira.inaf.it}\\
$1$ INAF - Istituto di Radioastronomia, via P. Gobetti 101, 40129 Bologna, Italy
}

\begin{document}

\date{Accepted 2013, November 27. Received ???; in original form ???}

\pagerange{\pageref{firstpage}--\pageref{lastpage}} \pubyear{2011}

\maketitle

\label{firstpage}

\begin{abstract}
Simulations of isolated binary mergers of galaxy clusters are a useful tool to study the evolution of these objects. For exceptionally massive systems they even represent the only viable way of simulation, because these are rare in typical cosmological simulations. We present a new practical model for these simulations based on the Hernquist dark matter profile. The hydrostatic equation is solved for a beta-model with $\beta = 2/3$ in this potential and approximate expressions for X-ray brightness and Compton-y parameter are derived. We show in detail how to setup such a system using SPH. The theoretical and several numerical models are compared to observed scaling relations of galaxy clusters and satisfactory agreement with the self-similar relations is found. The model is then applied to investigate the observed cluster ACT-CT J0102-4915 'El Gordo', a particularly massive merging high redshift cluster. We are able to reproduce the X-ray luminosity, SZ-effect and dark matter core distance as well as the rough shape of the observed cluster for reasonable model parameters. The lack of substructure prevents us from obtaining the fluctuations observed in the wake of the system and we argue that the parent cluster of the system was highly disturbed even before the main merger observed today.

\end{abstract}


\section{Introduction}\label{intro} 

In the Universe, matter is distributed in the form of a web-like structure. The largest knots in this cosmic web are galaxy clusters, which form the largest bound objects in the universe. The cluster gravitational potential is dominated by dark matter, in which baryonic matter settles to form the intra-cluster-medium (ICM).  Clusters are constantly subject to infall and merger processes by matter clumps of different size and mass, mostly along the filaments of the cosmic web. The largest of these events - major mergers - are the most energetic events in the universe, easily injecting $10^{63}\,\mathrm{erg}$ into the cluster volume. These processes are also connected to the thermal and non-thermal emission from clusters as they disturb and drive large scale shocks and turbulence in the ICM. The thermal emission of the ICM at $\approx 10^8 \,\mathrm{K}$ is observed in the X-ray band \citep[e.g.][]{2011A&A...535A...4R}. This as well as mass estimates from the Sunyaev-Zeldovich effect \citep{2009ApJ...701...32S} allows to confidently reconstruct the thermal properties of clusters. Furthermore a number of observed scaling relations imply approximate self-similarity of cluster \citep[e.g.][]{2011arXiv1112.5035B}.\par
Numerical simulations are well established as a way to study galaxy clusters using grid and particle codes \citep[e.g.][]{2008SSRv..134..229D,2010arXiv1005.1100N}. One usually simulates clusters as part of a larger (zoomed) cosmological simulation, which takes into account all the resolved substructure and its infall into the cluster. While this offers the greatest possible degree of realism, the approach is computationally costly and hard to analyse. In contrast direct simulations of two merging halos are more simplified as they do not contain any substructure. However these simulations are easy to analyse and computationally cheaper. Therefore they provide an ideal testbed for code evaluation and comparison. Additionally this approach can be used to model specific merging systems like the Bullet cluster and provide detailed insight into the dynamical state of the object.\par
Direct cluster merger simulations have been pioneered by \citet{1993ApJ...407L..53R} and \citet{1993A&A...272..137S}, who used the simulations to study the evolution of the X-ray brightness of merging clusters. The former group modelled a number of observed systems \citep{1995ApJ...446..583B,1998ApJ...493...62R,1999ApJ...518..603R},  the evolution of magnetic fields \citep{1999ApJ...518..594R} and subsequent non-thermal emission from radio relics in A3667 \citep{1999ApJ...518..603R}. The impact of mergers on cool-core clusters and cluster scaling relations has been studied by \citet{2002ApJ...569..122G,2006MNRAS.373..881P,2007MNRAS.380..437P,2008MNRAS.391.1163P}. \citet{2001ApJ...561..621R,2002MNRAS.329..675R,2007MNRAS.376..497M} presented parameter studies focusing on entropy production, and the generation of shocks and turbulence using SPH and grid methods.  \citet{2009ApJ...699.1004Z,2010ApJ...717..908Z} investigate idealised cool core clusters and core sloshing by merger interaction. They focused on the core of the cluster and the aspect of entropy generation and thermal conduction \citep{2011ApJ...728...54Z,2011ApJ...743...16Z}, as well as radio mini halos \citep{2013ApJ...762...69Z}. Recently \citet{2012MNRAS.419.1338R} investigated the impact of minor mergers of the generation of cold fronts and sloshing and on the generation of turbulence by Kelvin-Helmholtz instabilities \citep{2012MNRAS.420.3632R,2013ApJ...764...60R}. The impact of mergers on the magnetic field structure in a direct Eulerian MHD simulation was studied by \citet{2008ApJ...687..951T}. The off-set of X-ray and SZ peaks was studied by \citet{2012ApJ...748...45M}. DM self-interaction in clusters and the subsequent $\gamma$-ray signal has be investigated by \citep{2008ApJ...679.1173R}. \citet{2013MNRAS.429.3564D} investigated turbulent particle acceleration and the formation of radio halos in a direct cluster merger. \par
After the discovery of the Bullet cluster \citep{2002ApJ...567L..27M} direct simulations have been successfully used to model and investigate the observed system in detail, especially the kinematics, shock structure and temperature distribution \citep{2007ApJ...661L.131M,2007MNRAS.380..911S,2008MNRAS.389..967M,2012PASJ...64...12A}. Recently an  analysis of the merger kinematics of A1750 has been done by  \citet{2013arXiv1307.6643M} and of A3376 by \citet{2013MNRAS.430.3249M}.    \par 
Many previous studies consider an NFW profile for the DM distribution and a $\beta$-model for the gas distribution. The solution of the hydrostatic equation to obtain the gas temperature is usually done numerically. Alternatively some SPH models pre-relax the gas distribution from a constant temperature into hydrostatic equilibrium, letting the code obtain the temperature profile on its own (this usually results in a cool core cluster). \par
Regardless of the numerical algorithm, the description of the IC setup is often kept very concise concerning the technical details, especially in rather short studies. Therefore the large variety of techniques and models to setup cluster initial conditions make it hard to reproduce and compare between studies. The resulting setup is usually \emph{assumed} to roughly fit observed correlations. However the only works to actually show correlations are \citet{2007MNRAS.380..437P}  and \citet{2008MNRAS.389..967M}. Considering the literature it is difficult to identify an optimal model for the simulation of isolated galaxy clusters, suitable for cored as well as cuspy density profiles, providing long-term stable initial conditions and good match of the ICs to the observed scaling relations. In this paper we aim to fill this gap. Even though we focus on SPH here the model has been successfully implemented for the grid codes FLASH and ENZO as well (zuHone priv. comm.).\par
We present a model for initial conditions (IC) of direct cluster merger simulations using an analytical solution of the hydrostatic equation. We use a Hernquist dark matter density profile and a $\beta$-model with self-similar core radius and fixed slope for the gas density. We discuss the analytic predictions of the model, the numerical setup, the long term stability of the model, the merger setup and compare to the relevant cluster scaling relations. We also show how to extend it to cool-core clusters. For this model the hydrostatic equation can be solved analytically and X-ray brightness as well as SZ-effect can be approximated well. The resulting values roughly fit the main thermal scaling relations in the mass range of $10^{14}$ to a few $10^{15} \,M_\mathrm{sol}$. In addition the generation of initial conditions is comparatively easy and fast even at highest resolutions, due to the choice of DM profile and the analytic solution of the hydrostatic equation. We show how to sample a cluster merger beyond the virial radius to provide background gas and circumvent numerical problems with low viscosity formulations of SPH. \par
As a test case we then setup and simulate a cluster merger from a simple model of the cluster ACT-CL J0102-4915 ''El Gordo'' \citep{2012ApJ...748....7M}, including low viscosity gas. This cluster consists of a cool bullet and a disturbed system, colliding nearly head on. Due to its enormous size and high redshift a cluster similar to El Gordo is relatively unlikely to form in the standard cosmological model. It is therefore hard to find it in cosmological simulations, which cover only a limited volume (one would actually have to find a cluster with matching impact parameter, which is even harder). Furthermore the cluster has a particularly interesting Comet-like appearance in the X-ray band. The origin of this wake is still discussed. \par
This paper is organised as follows: In section \ref{model} we describe the analytical cluster model. In section \ref{implementation} we discuss details on the numerical implementation, and in \ref{scalings} we show a comparison with the thermal scaling relations and discuss the results. In section \ref{merger} we present a simple model simulation of the 'El Gordo' cluster. We draw our conclusions in section \ref{conclusions}.  \par

Throughout the paper we assume a concordance cosmology with $H_0 = 70 \, \mathrm{km}\, \mathrm{s}^{-1} \mathrm{Mpc}^{-1}$, $\Omega_{\mathrm{m},0}=0.3$, $\Omega_\Lambda$ = 0.7 and $\sigma_8=0.9$. This implies a critical density of $\rho_\mathrm{crit} = 9.2 \times 10^{-30} \, \mathrm{g} \,\mathrm{cm}^{-3}$ at $z=0$. 

\section{Cluster Model}\label{model}
Consider a spherically-symmetric cluster with mass $M_{200}$ with gas fraction $b_\mathrm{f}$.  Here $M_{200}$ is the mass enclosed inside  the radius $r_{200}$ so that the average density $\bar{\rho}(\left<r_{200})\right> = 200\, \rho_\mathrm{crit}$. For a given cluster mass in $r_{200}$, one can then calculate the normalisation of the cumulative mass density profiles, equation \ref{mod_DM_cum} and \ref{mod_cum_gas}. In what follows we set the gas fraction $b_\mathrm{f}(r_{500}) \approx 0.14$ (this corresponds to a gas fraction of 17\% in $r_{200}$), which is roughly consistent with observations and cosmological simulations \citep{2013MNRAS.431.1487P}. We do not consider cooling and star-formation. \par

\subsection{Dark Matter}\label{mod_DM}

{  Cosmological simulations show that the dark matter (DM) profile of galaxy clusters follows a near universal density profile. A number of profiles have been fit to the mass distributions found in numerical simulations \citep{1991ApJ...378..496D, 1996ApJ...462..563N, 2004MNRAS.349.1039N}, with the NFW profile being most widely used. Unfortunately it is still theoretically unclear why a near universal profile arises in collisionless DM clustering, so the exact analytic shape of the profile is not known \citep{2010gfe..book.....M}. \par
 While the NFW mass profile is most widely used, it does unfortunately not converge at large radii. For our model this would force us to introduce an artificial cut-off to the profile at large radii. Furthermore the NFW profile is rather impractical to use for particle sampling, as its distribution function is not analytically known. A convenient alternative to NFW models in isolated galaxy simulations was presented by \citet{2005ApJ...620L..79S}. They argue that the spherical Hernquist profile \citep{1990ApJ...356..359H,1991ApJ...378..496D,2002A&A...393..485B} is equal to the NFW profile at $r < 0.1 r_{200}$, but declines steeper thereafter ($\rho \propto r^{-2}$). Hence it converges at large radii, providing the required mass cut-off. As the central part of its potential is equal to the NFW potential we expect little effects on the dynamics of the two clusters.  Furthermore its distribution function is known, making it especially practical in numerical studies.  Alike to the NFW profile it represents a good fit to mass profiles inferred from observations of peculiar motions of galaxies in massive clusters \citep{2004AJ....128.1078R,2006AJ....132.1275R,2013A&A...558A...1B}. \par
We adopt the practical approach by \citet{2005ApJ...620L..79S} and model the DM component of the cluster as a Hernquist profile. Following cosmological simulations, \citet{2005ApJ...620L..79S} show that an NFW profile with concentration parameter $c$ corresponds to a Hernquist scale length $a$ , under the assumption of matching central densities:}
\begin{align}
    a = r_\mathrm{s} \sqrt{2(\ln(1+c) - c/(1+c)},
\end{align}
where $r_\mathrm{s} = r_{200}/c$.
The concentration parameter $c$ then varies with cluster mass as \citep{2008MNRAS.390L..64D}: 
\begin{align}
    c &= 5.74\left(\frac{M_{200}}{2\times10^{12} h^{-1} \,\mathrm{M_\mathrm{sol}}}\right)^{-0.097}
\end{align}

The DM density radial profile $\rho(r)$, the cumulative mass profile $M_\mathrm{DM}(<r)$ and the distribution function $f(E)$ are then given by \citep{1990ApJ...356..359H}:
\begin{align} 
    \rho_\mathrm{DM}(r) &= \frac{M_\mathrm{DM}}{2\pi}\frac{a}{r(r+a)^3} \\
    M_\mathrm{DM}(<r) &= M_\mathrm{DM} \frac{r^2}{(r+a)^2},\label{mod_DM_cum} \\ 
    f(E) &= \frac{M_\mathrm{DM}}{8\sqrt{2}\pi^3a^3v_\mathrm{g}^3} \left( 1-q^2 \right)^{-5/2} \left[ 3 \sin^{-1}\left(q\right) \right.\nonumber\\
     &+\left. q(1-q^2)^{1/2}(1-2q^2)(8q^4 - 8q^2 - 3)\right],\label{mod_DM_dfunc}
\end{align}
where
\begin{align}
    v_\mathrm{g} &= \sqrt{\frac{G M_\mathrm{DM}}{a}}, &     q &= \sqrt{-\frac{E}{v_\mathrm{g}}}. \nonumber
\end{align}

\subsection{Baryonic Matter}\label{mod_bm}

Motivated from observations \citep[e.g.][]{2008A&A...487..431C} we assume the radial gas distribution of the ICM to follow the well known beta-model \citep{1978A&A....70..677C}:
\begin{align}
    \rho_{\mathrm{gas}}(r) &= \rho_0 \left( 1 + \frac{r^2}{r^2_{\mathrm{c}}} \right)^{-\frac{3}{2}\beta}, \label{mod_gas}
\end{align}
where $\rho_0$ is the central gas density and we set \citep{2008MNRAS.389..967M}: $\beta = \frac{2}{3}$. To model a disturbed cluster we assume $r_{\mathrm{c}} = \frac{1}{3}r_{\mathrm{s}}$, to model a cool core cluster we set $r_{\mathrm{c}} = \frac{1}{20}r_{\mathrm{s}}$. For this choice we find the cumulative radial gas mass distribution: 
\begin{align}
    M_\mathrm{gas}(<r) &= 4\pi r_\mathrm{c}^3 \rho_0 \left[ \frac{r}{r_\mathrm{c}} - \arctan\left(\frac{r}{r_\mathrm{c}}\right) \right]. \label{mod_cum_gas}
\end{align}
Note that for other choices of $\beta$ this step involves the confluent hypergeometric function, which usually makes the analytical solution of the hydrostatic equation impossible. Another elegant choice would be $\beta=1$, which however is unfavoured from observations.\par
\begin{figure*}
\includegraphics[width=0.49\textwidth]{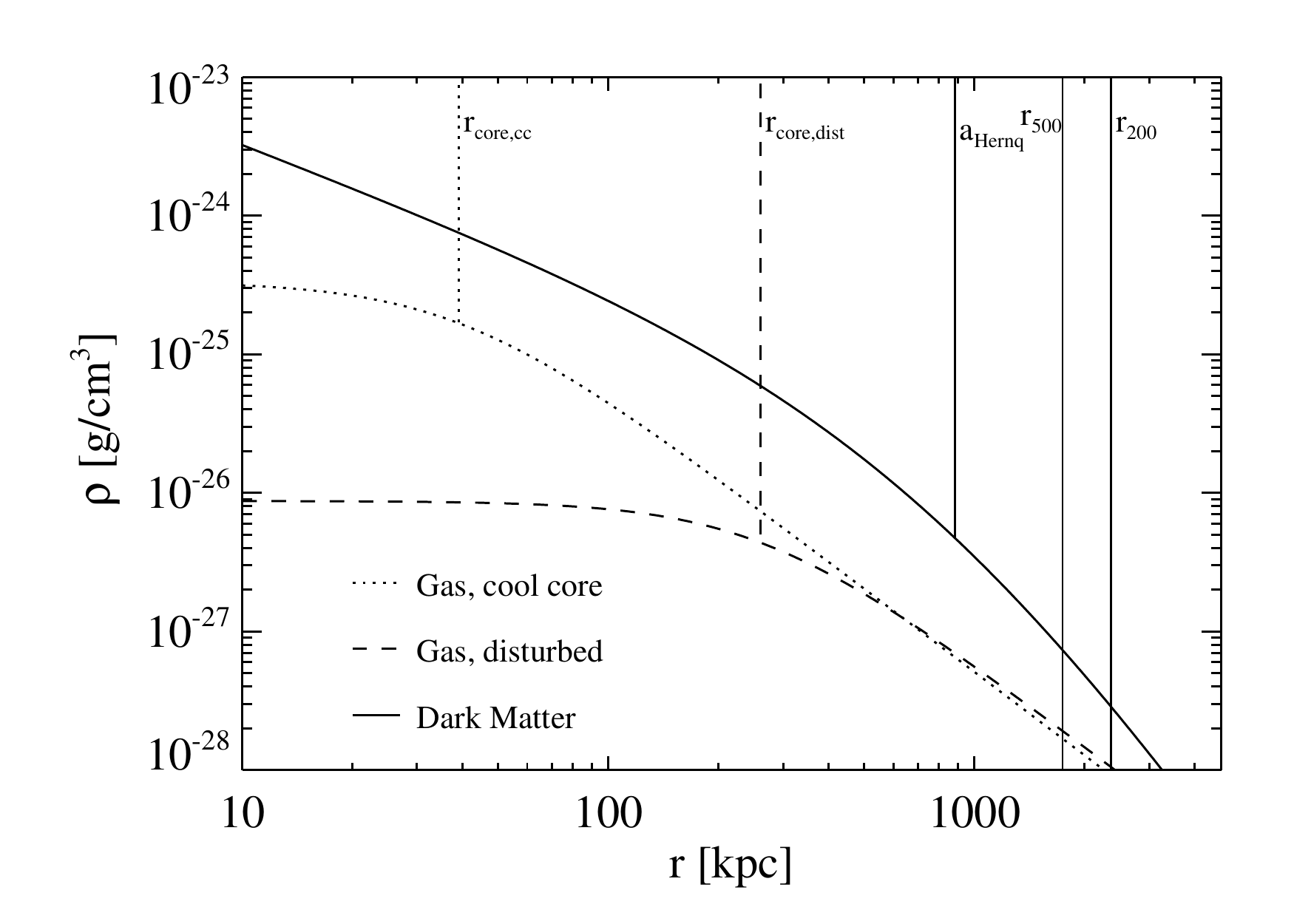}
\includegraphics[width=0.49\textwidth]{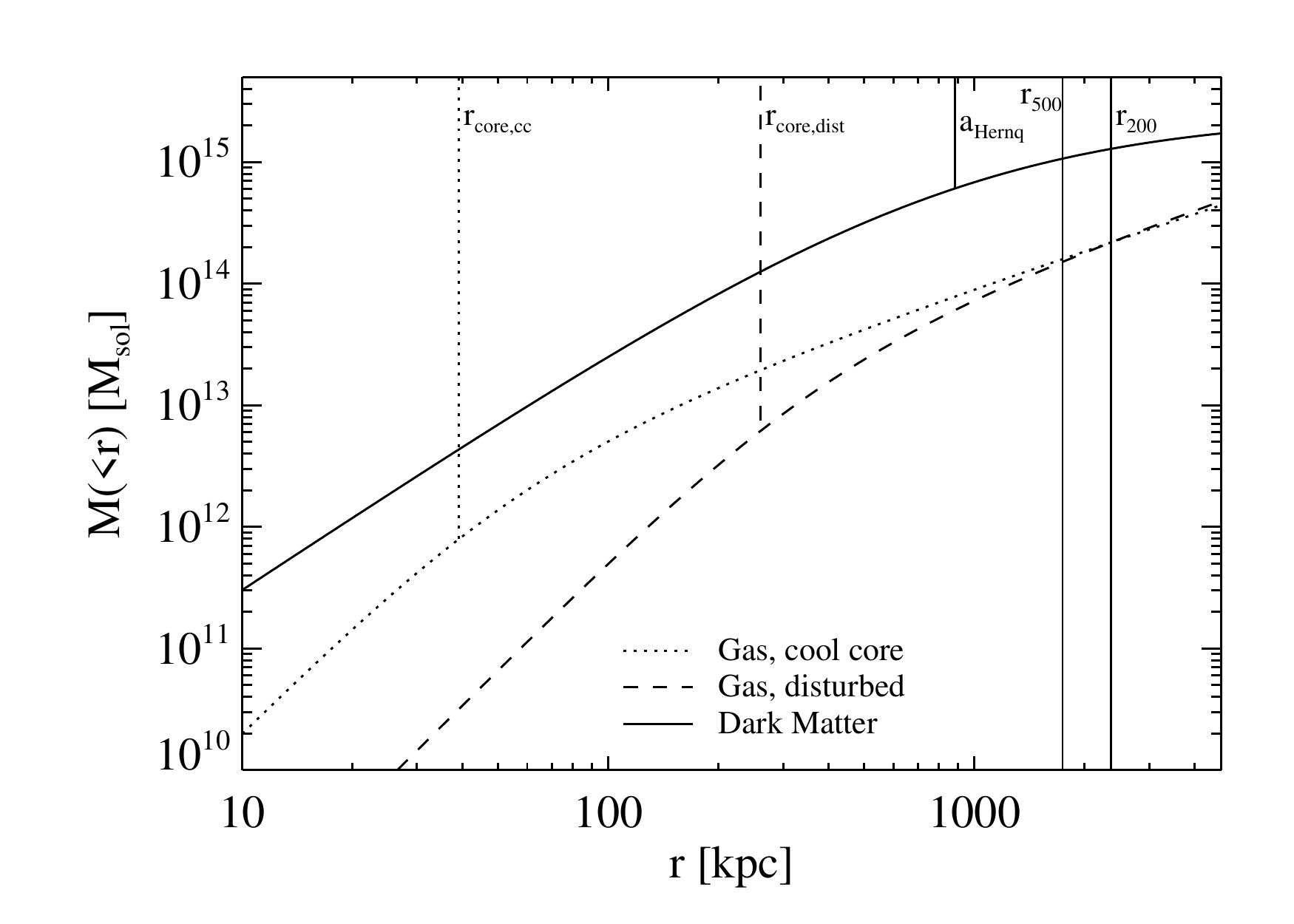}
\includegraphics[width=0.49\textwidth]{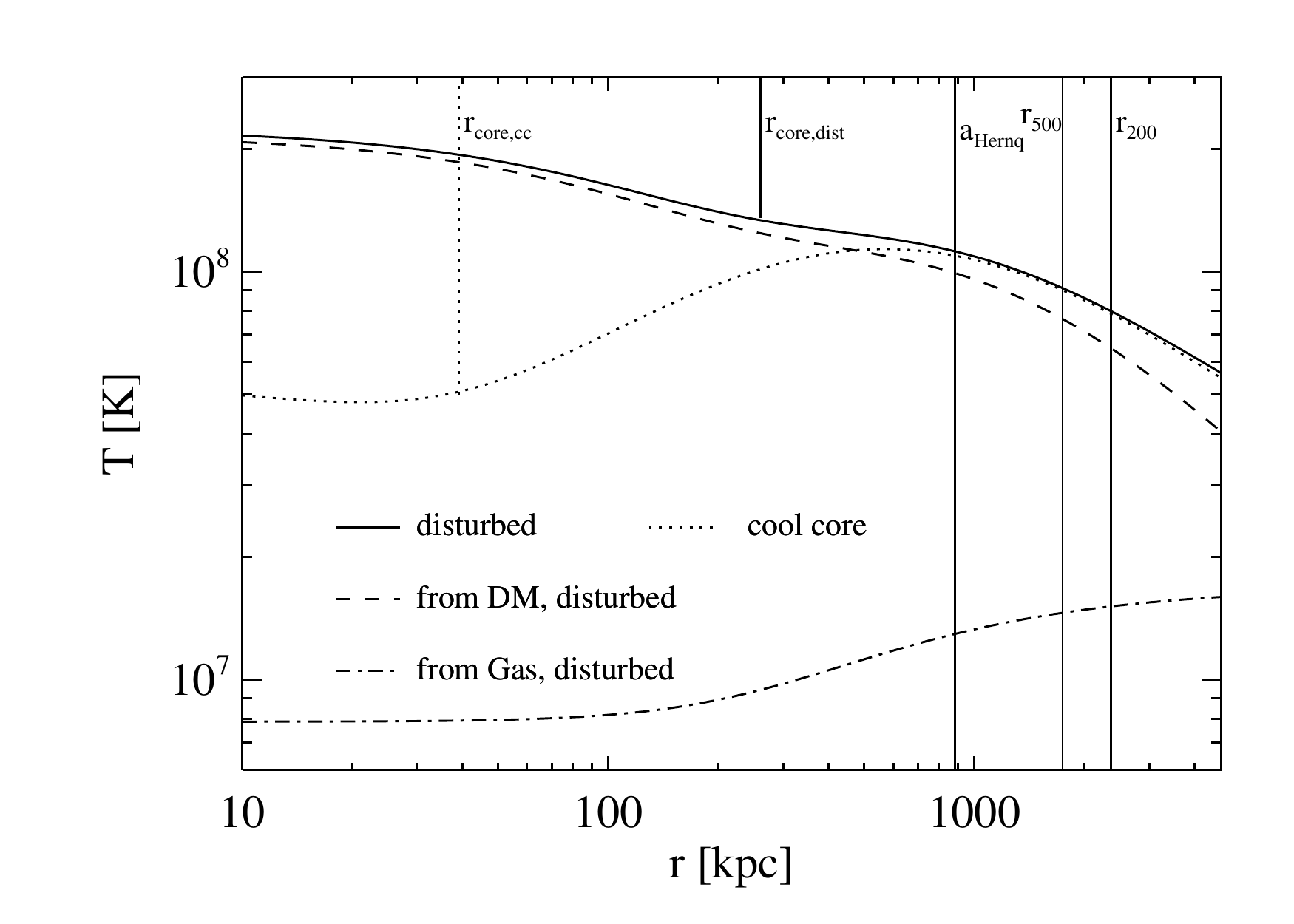}
\includegraphics[width=0.49\textwidth]{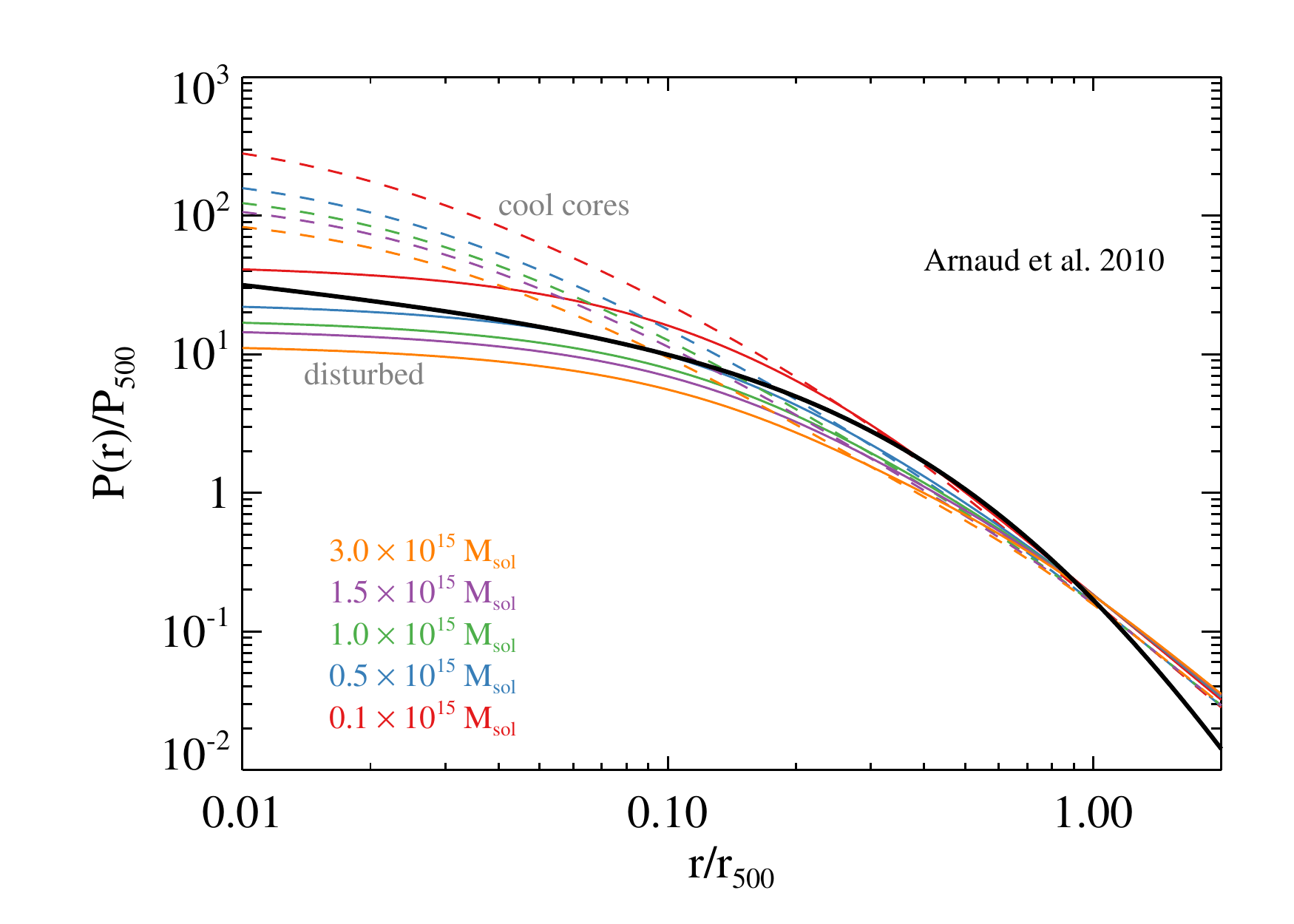}
\caption{Radial profiles for a cluster with $M_{200} = 1.5 \times 10^{15} \, M_\mathrm{sol}$, disturbed (dashed lines) and as cool core (dotted lines). From top left to bottom left: DM and gas density, cumulative mass, and temperature. On the bottom right we show universal pressure profiles from the model of disturbed (full lines) and cool core clusters (dashed) with masses between $M_{200} =  10^{14} \, M_\mathrm{sol}$, and $M_{200} = 3.0 \times 10^{15} \, M_\mathrm{sol}$ in colours. We over-plot the universal pressure profile from \citet{2010A&A...517A..92A} as black line.}\label{img_model}
\end{figure*}

In a relaxed galaxy cluster the intra-cluster-medium is in approximate hydrostatic equilibrium in the gravitational potential of the cluster:
\begin{align}\label{mod_bm_he}
    \frac{1}{\rho_\mathrm{gas}} \frac{\mathrm{d}P_\mathrm{gas}}{\mathrm{d}r} &= -G\frac{M_\mathrm{total}(<r)}{r^2},
\end{align}
so that with the ideal gas law: $P = nk_{\mathrm{B}}T$, the temperature of the ICM in hydrostatic equilibrium is given by \citep[e.g.][]{2005MNRAS.363..509M}: 
\begin{align}
    T(r) &= \frac{\mu m_{\mathrm{p}}}{k_{\mathrm{B}}} \frac{G}{\rho_{\mathrm{gas}}(r)} 
        \int\limits^{R_{\mathrm{max}}}_{r} \frac{\rho_{\mathrm{gas}}(t)}{t^2} M_\mathrm{total}(<t) \,\mathrm{d}t ,
\end{align}
where $R_{\mathrm{max}}$ is the maximum sampling radius. This can be solved   given eq. \ref{mod_DM_cum} and eq.\ref{mod_cum_gas} and for $R_{\mathrm{max}} \rightarrow \infty$: 
\begin{align}
    T(r) &=  G \frac{ \mu m_{\mathrm{p}}}{k_{\mathrm{B}} }  \, \left(1+\frac{r^2}{r^2_{\mathrm{c}}}\right) \left[ M_\mathrm{DM}  F_0(r) + 4\pi r_\mathrm{c}^3 \rho_0 F_1(r) \right]\label{mod_temp}
\end{align}
where we define: 
\begin{align}
    F_0(r) &= \frac{r_\mathrm{c}}{(a^2+r_\mathrm{c}^2)^2} \left[ \frac{\pi}{2}(a^2-r_c^2) +r_\mathrm{c} \frac{a^2+r_\mathrm{c}^2}{a+r}      \right. \nonumber\\ 
    &- \left. \left(a^2 - r_\mathrm{c}^2\right) \arctan\left(\frac{r}{r_\mathrm{c}}\right) -  r_\mathrm{c} a \ln\left(\frac{(a+r)^2}{r^2+r_\mathrm{c}^2} \right)   \right] \\
    F_1(r) &= \frac{\pi^2}{8r_\mathrm{c}} -\frac{\arctan^2 \left(r/r_\mathrm{c}\right)}{2r_\mathrm{c}} -\frac{\arctan(r/r_\mathrm{c})}{r}. 
\end{align}
It is helpful to define the characteristic temperature of the cluster $T_c = T(r_c)$, which is then given by :
\begin{align}
    T_c &= 2 G \frac{\mu m_\mathrm{p}}{k_\mathrm{B}} \left[ M_\mathrm{DM} \frac{r_\mathrm{c}^2}{(a^2 + r_\mathrm{c}^2)^2} \times \right. \nonumber\\ 
    &\times \left( \frac{\pi}{4r_\mathrm{c}}(a^2 - r_\mathrm{c}^2) + \frac{a^2 + r_\mathrm{c}^2}{a+r_\mathrm{c}} - a\ln\left(\frac{(a+r_\mathrm{c})^2}{2r_\mathrm{c}^2}\right)\right) \nonumber\\
    &+ \left. \pi^2 r_\mathrm{c}^2 \rho_0 \left( \frac{3\pi}{8}-1 \right) \right] \label{eq.Tc}
\end{align}
In figure \ref{img_model}, first three panels,  we plot model properties for a cluster with mass of $M_{200} = 1.5 \times 10^{15} \, \mathrm{M}_\odot$ with $R_\mathrm{max} \rightarrow \infty$. We show gas profiles for a cool core cluster (dotted) and a disturbed cluster (dashed). DM and gas density in the upper left panel; DM, gas and total cumulative mass in the upper right; gas temperature in the lower left. Here we show the disturbed cluster profile as full line. We also show the contributions of the gas and DM potential to the temperature as dashed and dashed-dotted line, respectively. We mark the Hernquist scale length, $r_{200}$ and $r_{500}$ as vertical line, and the ICM core radius as dotted line for a cool core cluster at $\approx 35 \, \mathrm{kpc}$ and as dashed line for a disturbed cluster at $\approx 250 \,\mathrm{kpc}$. At the bottom right we show a comparison of the pressure profile of a number of disturbed (full lines) and cool core (dashed) models to the observed universal pressure profile from \citet{2010A&A...517A..92A}.  Here we obtain an acceptable fit inside $r_{500}$ for our model. Deviations from the self-similar fit are consistent with the adiabatic models in \citet{2007ApJ...668....1N}, i.e. arise, because we do not include cooling and star formation.

\subsection{Observables}

The bolometric X-ray luminosity of the ICM can be approximated  by the free-free Bremsstrahlung emission of the electrons in the plasma \citep[e.g.][]{1986rpa..book.....R}:
\begin{align}
    L_\mathrm{X} &= \sqrt{\frac{2\pi k_\mathrm{B}}{3 m_\mathrm{e}}} \frac{2^5 \pi e^6}{3 h m_\mathrm{e} c^3 } \int \limits_{V} \sqrt{T} n_\mathrm{th}^2 \bar{g}_\mathrm{B} \, \mathrm{dV} 
\end{align}
where $\bar{g}_\mathrm{B} \approx 1.2$ is the average Gaunt factor, and $n_\mathrm{th}$ is the number density of ions and electrons. Because the bulk of the X-ray emission is located at the core radius of the cluster, we set $T(r) = T_c$ so the integral becomes analytically treatable. We find then from equations \ref{mod_temp} and \ref{mod_gas}:
\begin{align}
    L_\mathrm{X, model} &=  \sqrt{\frac{2\pi k_\mathrm{B}}{3 m_\mathrm{e}}} \frac{2^5 \pi e^6}{3 h m_\mathrm{e} c^3 } \sqrt{T(r_c)} \left( \frac{\rho_0}{\mu m_\mathrm{p}} \right)^2 \nonumber\\
    &\times 4\pi r_c^3 \left[ \arctan\left(\frac{r}{r_c}\right) - \frac{r r_c}{r_c^2 + r^2} \right]^{R_\mathrm{max}}_0 \label{eq.xray} \\
    &\approx 2.2\times 10^{-27} r_\mathrm{c}^3 \sqrt{T_\mathrm{c}} \left( \frac{\rho_0}{\mu m_\mathrm{p}} \right)^2 \nonumber
    \end{align}
The volume integrated Compton-y parameter is given by \citep{1980ARA&A..18..537S} :
\begin{align}
    y &= \frac{\sigma_\mathrm{T}k_\mathrm{B}}{m_\mathrm{e} c^2} 4 \pi  \int\limits_{0}^{R_\mathrm{max}} r^2 n_e(r) T(r) \,\mathrm{d}r.
\end{align}
Because observationally the Compton-y parameter is usually quoted inside $r_{500}$ and the temperature is roughly constant, we set $T(r) \approx T(r_{500})$ and get from \ref{mod_temp} and \ref{mod_gas}:
\begin{align}
	y_\mathrm{model}(R_\mathrm{max}) &= 4 \pi r_\mathrm{c}^2 T(r_{500}) \frac{\sigma_\mathrm{T}k_\mathrm{B}}{m_\mathrm{e} c^2} \frac{\rho_0}{\mu m_\mathrm{p}} F_\mathrm{y}(R_\mathrm{max}) \label{eq.sz} \\
								 & \approx 1.41\times 10^{-33}  r_\mathrm{c}^2 T(r_{500}) \frac{\rho_0}{\mu m_\mathrm{p}} F_\mathrm{y}(R_\mathrm{max}) \nonumber\\
	F_\mathrm{y}(R_\mathrm{max}) &=  \left[ R_\mathrm{max} - r_\mathrm{c} \arctan\left(\frac{R_\mathrm{max}}{r_\mathrm{c}}\right)\right] 
\end{align}
This diverges for $R_\mathrm{max} \rightarrow \infty$ \citep[see also][]{2007ApJ...665..911H,2010A&A...517A..92A}. However from observations we are usually interested in the SZ-signal inside $r_{500}$ or $r_{200}$, for which the estimate is not problematic.  \par
The predicted correlations from the equations \ref{eq.Tc}, \ref{eq.xray} and \ref{eq.sz} are shown and discussed in section \ref{scalings}.

\section{Implementation} \label{implementation}
We implemented the analytical model in a C code to produce initial conditions for the MHD-SPH code {\small GADGET-3} \citep{springel2005,2009MNRAS.398.1678D}. All simulations are run using the Wendland C6 kernel \citep{2012MNRAS.425.1068D} with a compact support $h_\mathrm{sml}$ ('smoothing length') so that the kernel weighted number of particles inside that scale is $295 \pm 0.01$. We use a gravitational softening of a seventh of the  length that corresponds to the mean volume per particle in the box. \par
We split the particle number equally between SPH and DM particles. The particle mass can then be calculated from $M_{200}$ using equations \ref{mod_DM_cum} and \ref{mod_cum_gas}. Our choice of a Hernquist profile for the DM distribution has the advantage to be easily invertible, so we can sample the mass profile via:
\begin{align}
    q &= \frac{M_\mathrm{DM}(<r)}{M_\mathrm{DM}}\\
    \Rightarrow r &=  \frac{a\sqrt{q}}{1-\sqrt{q}}, \label{imp_smap_DM}
\end{align}
where $q$ is a random number between zero and one. As equation \ref{mod_cum_gas} can not be inverted analytically, we sample the gas mass by numerically inverting it through a table. \par
The velocity of the DM component is initialised so the Hernquist distribution function (DF),  equation \ref{mod_DM_dfunc}, is satisfied. Here we are using the rejection sampling algorithm \citep{vonNeumann:1951:VTU,1992nrfa.book.....P}. For every particle we first calculate the maximum potential energy it may have, given it's position. We then calculate the maximum of the DF for that particle. Then we draw a random velocity and calculate the value of the DF for that velocity. Only if this value is smaller than another random number times the maximum of the DF the velocity is accepted. \par
The initial velocity of the gas particles is set to zero. The gas temperature follows straight forward from equation \ref{mod_temp}.\par
The clusters DM profile is sampled up to infinity. The gas profile of one cluster is sampled up to a maximal radius $R_\mathrm{max}$, which is half the box size. This provides a natural way to embed the observable inner region of the cluster in a background density and keep boundary effects away from the cluster centre. Note however that for large $R_\mathrm{max}$ the baryon fraction gets larger than one in the outer cluster regions. This is not a problem because at these radii our spherical symmetric model has limited validity anyway.

\subsection{Relaxation \& Stability}
\begin{figure*}
\includegraphics[width=0.49\textwidth]{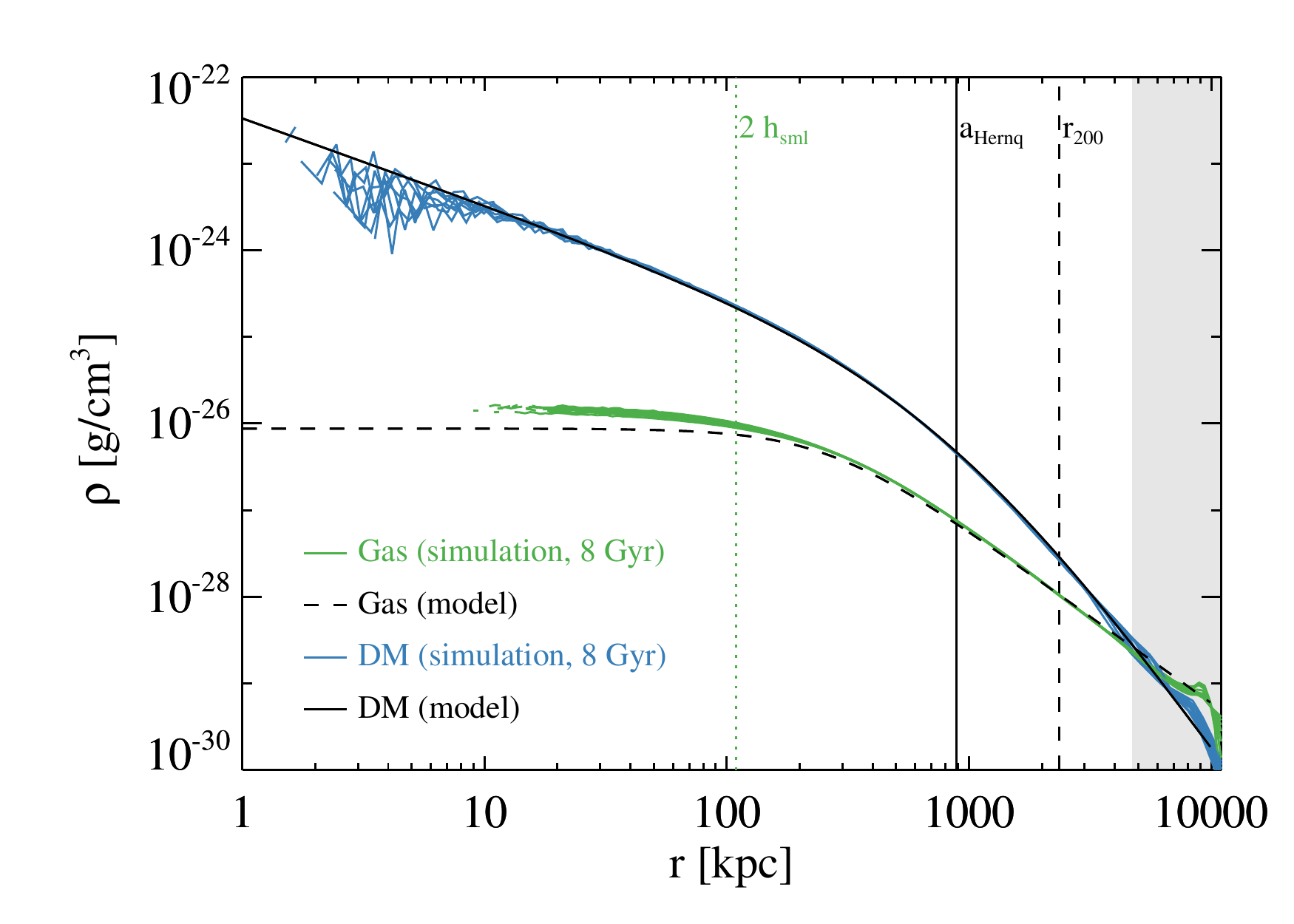}
\includegraphics[width=0.49\textwidth]{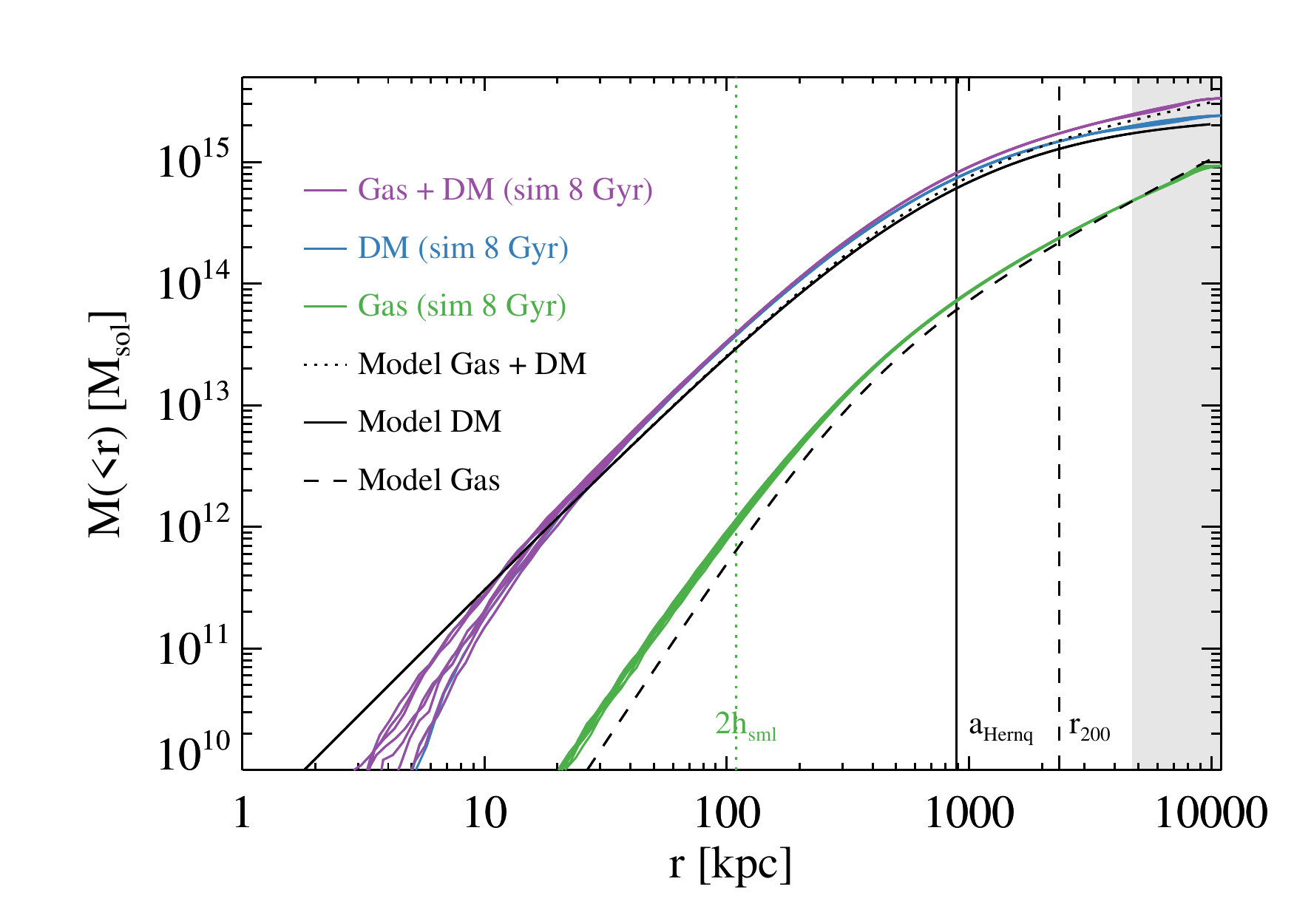}
\includegraphics[width=0.49\textwidth]{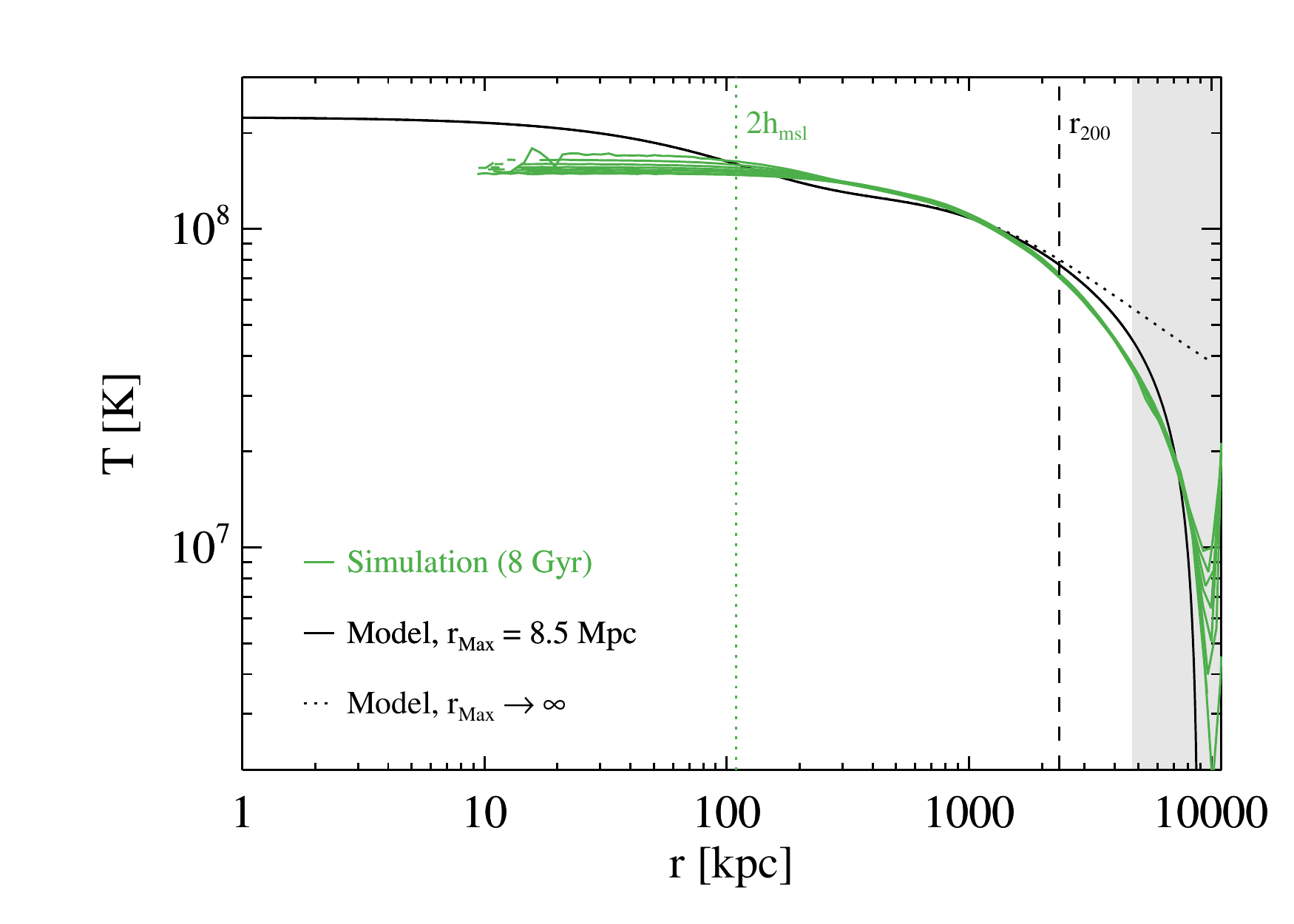}
\includegraphics[width=0.49\textwidth]{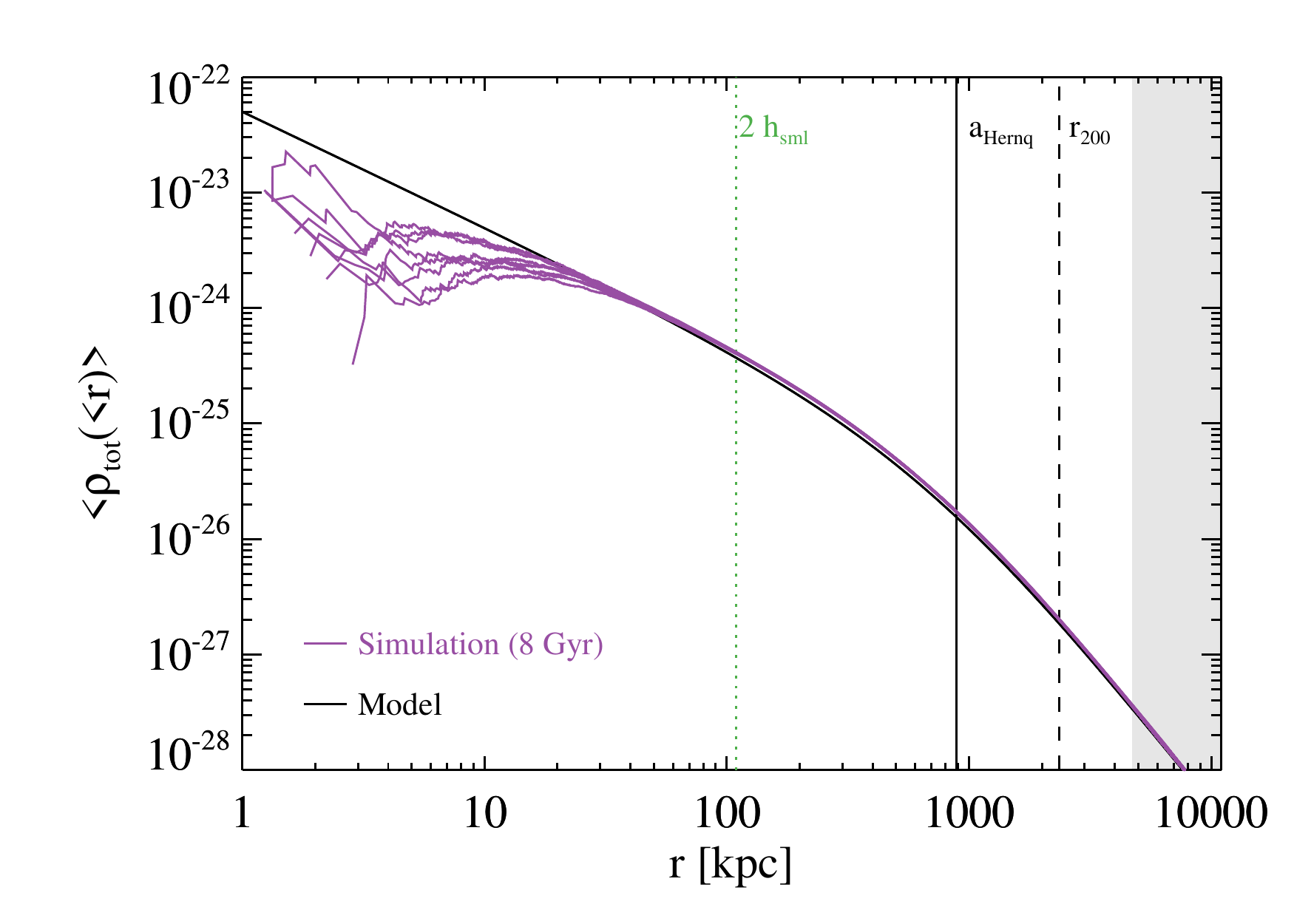}
\caption{Analytical (black lines) and numerical model (colours) for a cluster with $M_{200} = 1.5 \times 10^{15}\, M_\mathrm{sol}$. The numerical model was evolved for 8 Gyr to demonstrate stability. We show simulated DM in blue, gas in green and combined profiles in purple. From top left to bottom right: density, cumulative mass, temperature and mean total density over radius. We mark the resolution scale as vertical green dotted line, the Hernquist scale length as vertical black line and $r_{200}$ as vertical dashed line. In the temperature plot we show the temperature for the numerical model including surface terms as full line and without surface terms (i.e. $r_\mathrm{max} \rightarrow \infty$). The boundary region, where the model is not valid, is shaded grey. }\label{img_numodel}
\end{figure*}
The Poisson sampling of the gas density profiles introduces fluctuations into the SPH density estimate, as this does not result in a minimum energy state of the particles \citep{2012MNRAS.425.1068D}. These density fluctuations translate into temperature and velocity fluctuations, when the simulation is started, i.e. sub-kernel scale noise. If a low-viscosity scheme is used, these fluctuations will cause unwanted viscosity all over the simulation domain and reduce the overall Reynolds number of the flow. In addition the inhomogeneous sampling implies large error terms \citep{2012JCoPh.231..759P} and cause large gradient errors in the simulation \citep[e.g.][]{2012MNRAS.423.2558B}. \par
To resolve this issue, we pre-relax the cluster in a periodic box for a couple of Gyr. We use the low-viscosity scheme implementation of \citet{2005MNRAS.364..753D} and start with a very large viscosity parameter $\alpha=25$, a large decay length $l = 4$, a source scaling of $S_i = 6$ and a minimum parameter of $\alpha_\mathrm{min} = 0.02$. To equilibrate fluctuations in temperature we use the artificial thermal conduction by \citet{2008JCoPh.22710040P}, with a large conduction parameter of 4. This slowly relaxes the particles into a low energy state and ensures that the viscosity parameter remains low outside of shocks later. The corresponding particle sampling of temperature and density is homogeneous, i.e. without fluctuations on scales of the kernel FWHM. Note that any SPH simulation without this property is broken and should be discarded \citep{2012ASPC..453..249P} \par 
To test our implementation for numerical stability we evolve a relaxed cluster with $M_{200} = 1.5 \times 10^{15} \, \mathrm{M}_\mathrm{sol}$, sampled by $5 \times 10^{6}$ particles up to $R_\mathrm{max}  = 8.5 \, \mathrm{Mpc}$ for 6 Gyrs. This implies an effective total mass of $M_\mathrm{tot} = 3.2 \times 10^{15}\, \mathrm{M}_\mathrm{sol}$. \par
In figure \ref{img_numodel} we plot the evolution of DM (gas) density, cumulative mass, temperature and average total density, respectively. The gas is shown in green, the DM in blue, combined profiles in violet, the analytical model in black. We mark the gas resolution scale (vertical dotted line), the Hernquist scale (full line) and $r_{200}$ (dashed line). There is only little evolution in the gas density profile of the cluster and a slight variation in temperature over the 6 Gyrs of evolution. The inner part of the model suffers from resolution effects, i.e. the DM density is noisy and the SPH density and temperature estimates are smoothed, as expected. In the simulation the inner temperature is lower compared to the model, so a small excess of gas density is present in the cluster centre.  The additional DM mass in the centre of the box (see section \ref{sect.box}) does not affect the density profile much. \par

\subsection{Boundary Regions and the Periodic Box}\label{sect.box}
We embed the cluster into a periodic box of size $L_\mathrm{Box} = 3.75 R_{200}$. This choice has the advantage that: 
\begin{enumerate}
    \item stray SPH particles can not escape the cluster region. In case of a merger this causes low timesteps for these particles and is bound to happen, when a low-viscosity scheme is used.
    \item the cluster has only a small boundary region in the corners of the box. Subsequent changes to the density and temperature profiles at the cluster edge are minimised.
    \item the specific choice of the box size ensures that the slope of the Y-M correlation is correct.
\end{enumerate}
However, as the DM profile is sampled to infinity, the code will map all particles inside the box in the beginning. This means that more DM particles will be in the center of the cluster than predicted by the profile. These belong to the 'neighbouring' clusters and decrease the effective baryon fraction in the central region. This alters the cumulative mass profile (image \ref{img_numodel}, top right) and increases $M_{500}$ in the simulation. This affects $M_{500}$ in the scaling relations (figure \ref{img_scaling}). For small box sizes the stability of the cluster would suffer as well.  However for large enough boxsizes stability is unproblematic. \par
At the boundary regions our cluster model is not a physical model anymore. The gas density becomes higher than the DM density in the model. Furthermore the finite sampling radius implies a drop in temperature (surface terms), which start to become significant beyond $2 r_{200}$. At these radii, real clusters are dominated by filamentary accretion and the subsequent accretion shock and turbulence. Nonetheless the cold gas at large radii provides a background density for our simulation and the large boxsize keeps boundary effects away from the interesting cluster centre. As a consequence, some evolution of the density can be seen beyond $2r_{200}$. Here the edge of the periodic box and then 'next cluster' are also leading to an increase in temperature. \par
Note that considerable effort is necessary to minimize numerical errors in the centre of the cluster and push boundary effects out of the physically interesting region ($r_{200}$). This is especially true for SPH models. Numerical models for cluster mergers usually focus on the inner regions of the objects, which are unaffected by the boundaries, if the clusters are stable.

\section{Scaling Relations}\label{scalings}
\begin{figure*}
    \includegraphics[width=0.49\textwidth]{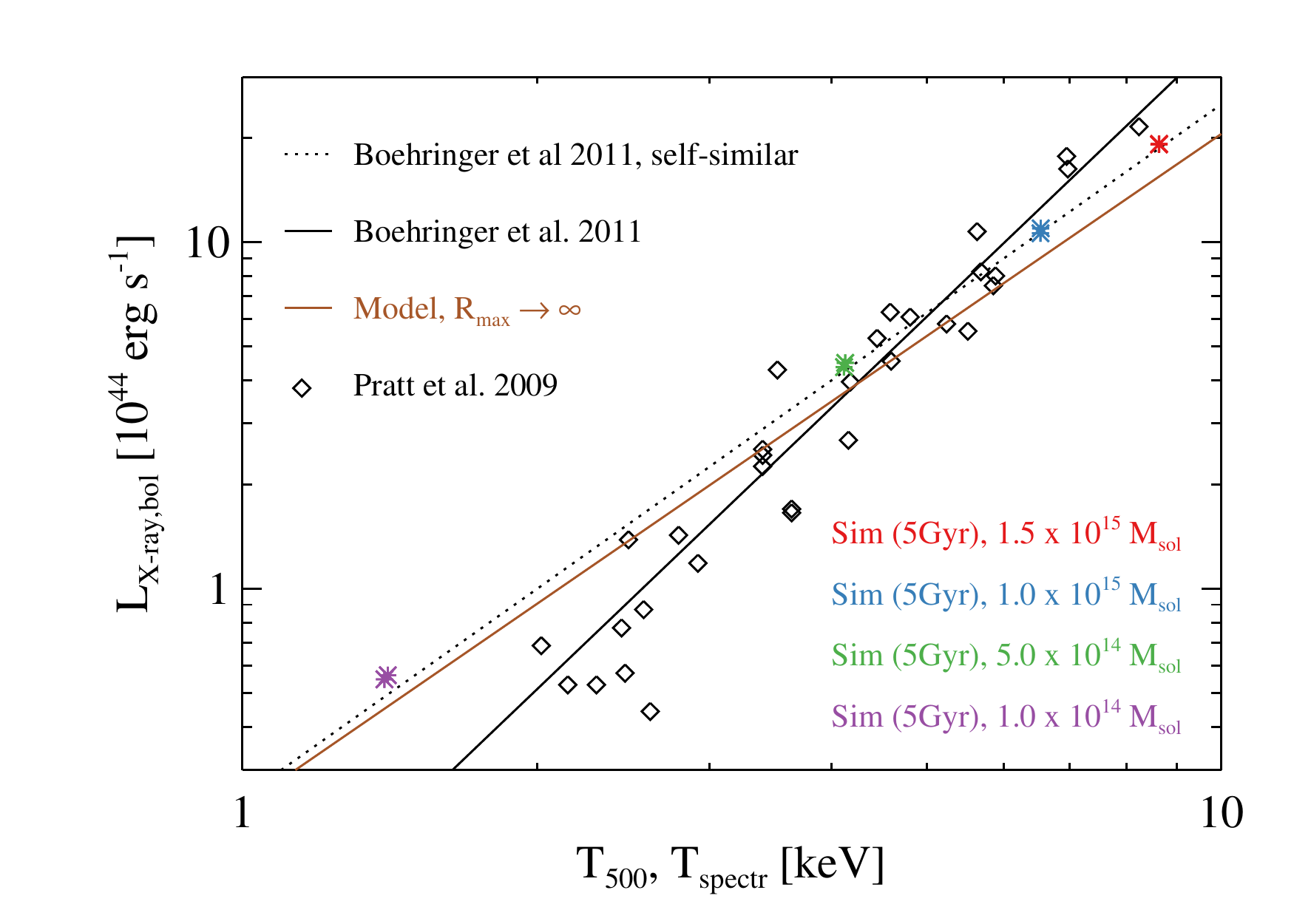}
    \includegraphics[width=0.49\textwidth]{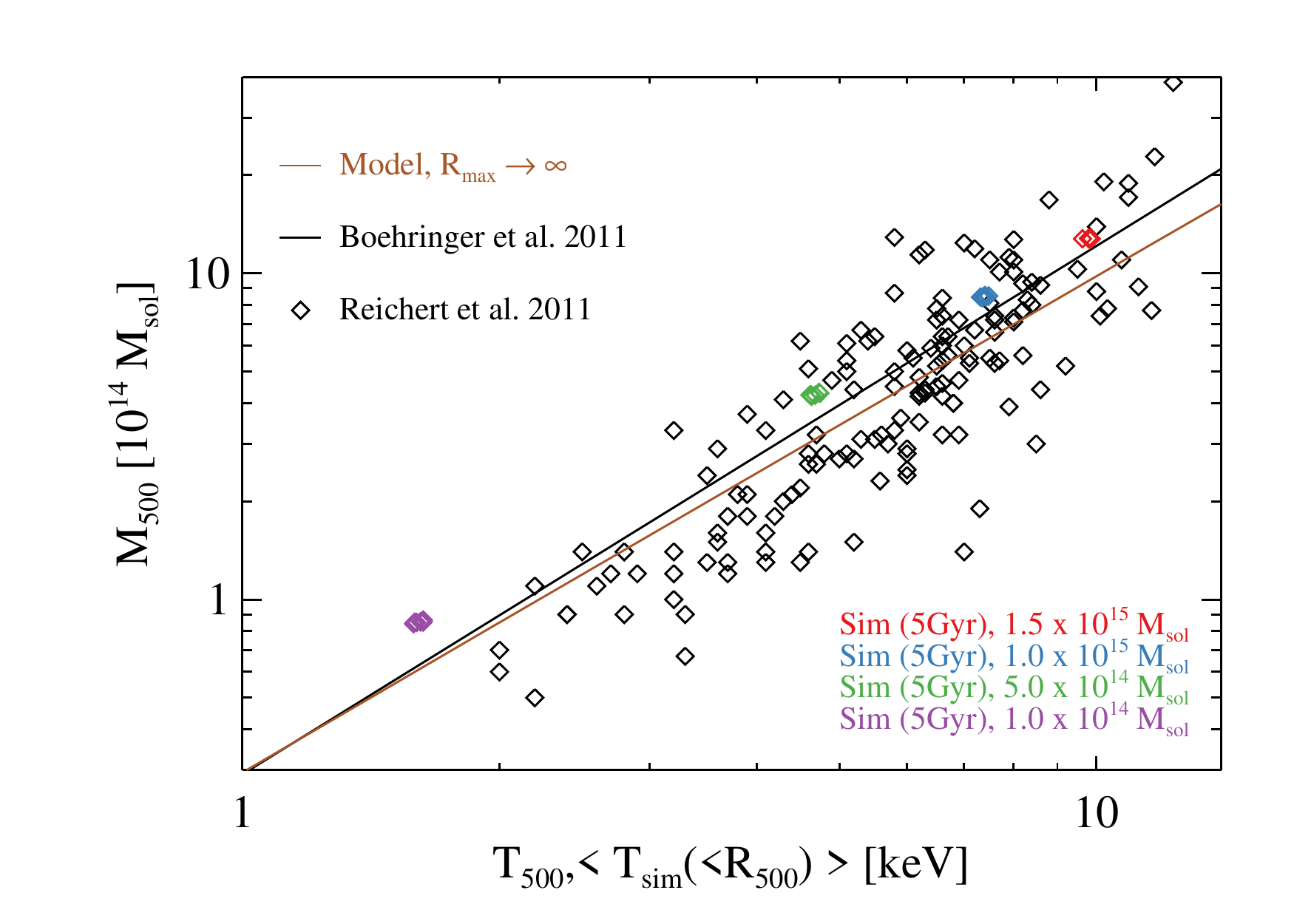}
    \includegraphics[width=0.49\textwidth]{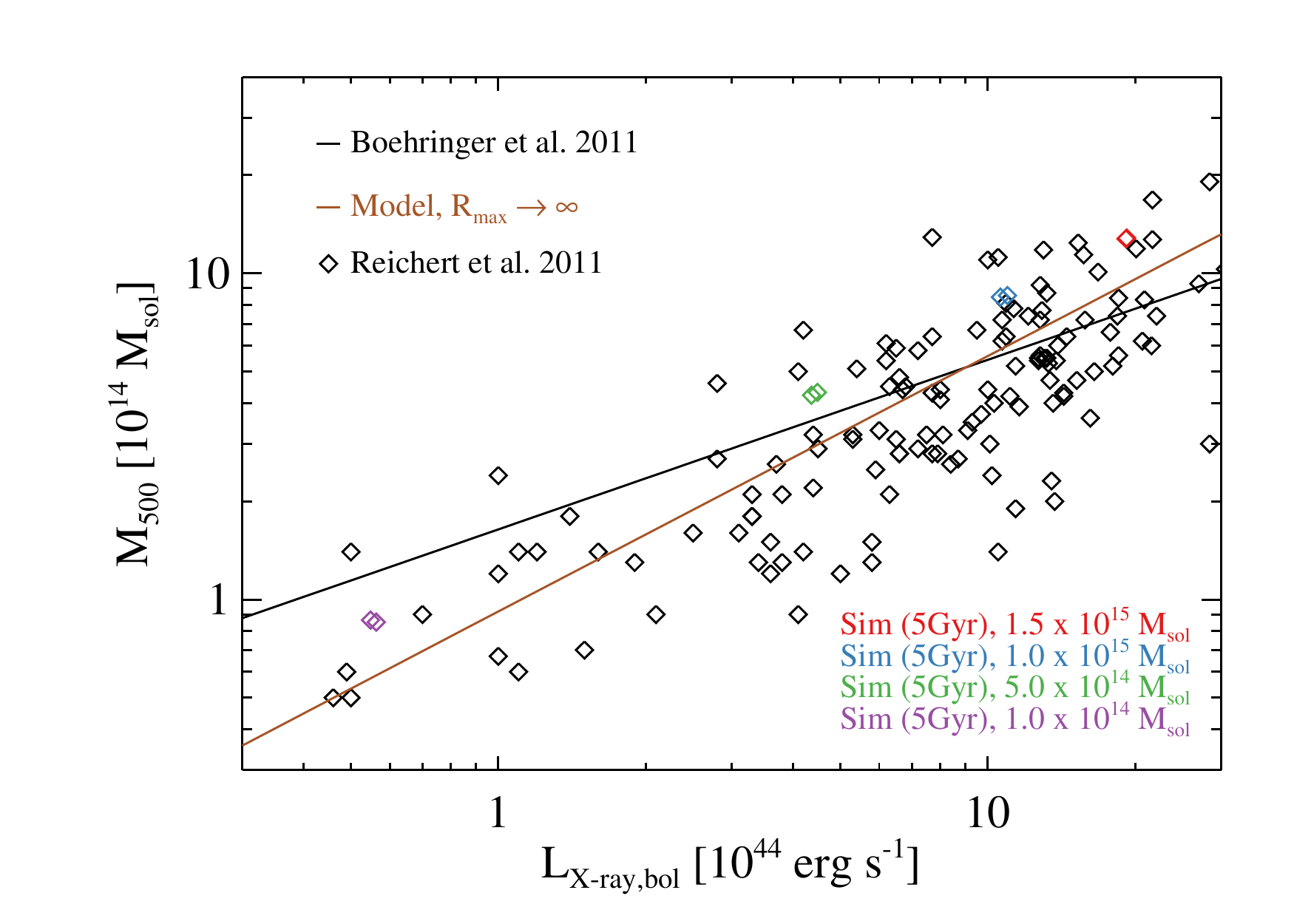}
    \includegraphics[width=0.49\textwidth]{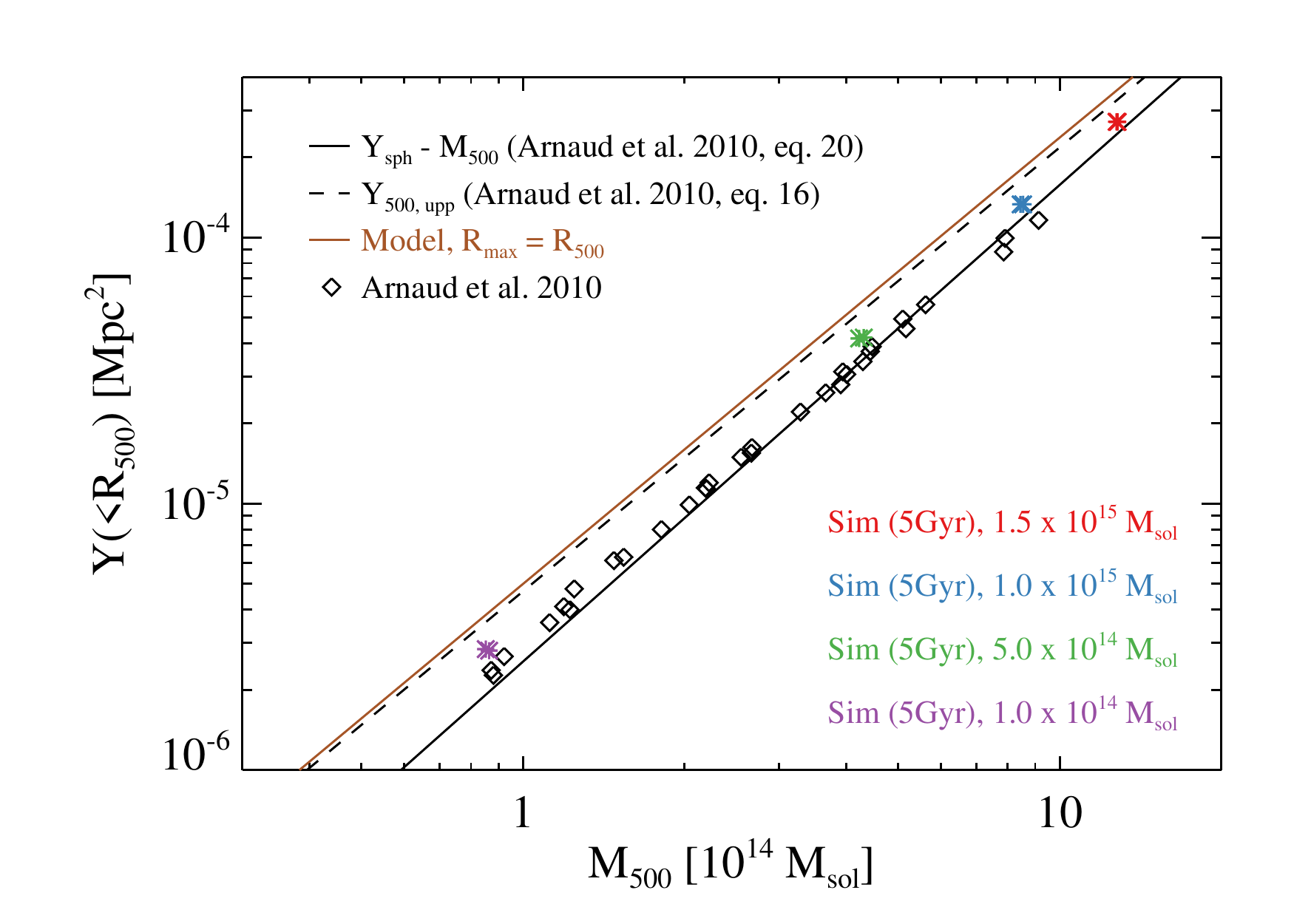}
    \caption{Top left: Bolometric X-ray luminosity over temperature in $r_{500}$, observations of individual clusters as black diamonds from \citet{2009A&A...498..361P}, observed (black line) and self-similar (dotted line) correlation from \citet{2011arXiv1112.5035B}, our model (eq. \ref{eq.xray} and \ref{eq.Tc}) in brown, simulated cluster as coloured diamonds where we take the spectroscopic temperature estimate in $r_{500}$ \citep{2004MNRAS.354...10M}. Top right: Observed M-T relations in $r_{500}$ as black line \citep{2011arXiv1112.5035B}, observed clusters as black diamonds \citep{2011A&A...535A...4R}, our model in brown and simulated clusters in colours, where we take $M_{500}$ and $T_{500}$ directly from the simulation. Bottom left: Observed $M-L_\mathrm{x}$ relation in black \citep{2011arXiv1112.5035B} and individual clusters as black diamonds \citep{2011A&A...535A...4R}, our model in brown and simulated cluster as coloured diamonds. Bottom right: Integrated Compton-Y parameter in $r_{500}$ over $M_{500}$, observed correlation and individual clusters in black (diamonds) from \citep{2010A&A...517A..92A}, our model in brown, simulated clusters in colours. We also add the characteristic Compton-Y parameter at $r_{500}$ from the universal pressure profile as dashed line from \citet{2010A&A...517A..92A}.}\label{img_scaling}
\end{figure*}

Structure formation and subsequently the self-similar model of DM halos \citep{1974ApJ...187..425P} predict a number of scaling relations for global cluster properties like mass, temperature and X-ray luminosity \citep[e.g.][]{2011arXiv1112.5035B}.
{ We compare the predictions of our analytical and numerical model to the observed relations. While we do not expect surprises here, we would like to understand:
\begin{itemize}
    \item How different is our analytical model from actual observed clusters. I.e. how large is the scatter in the observations compared to the model.
    \item Does the numerical model agree well with the expectations from theory. I.e. does the relaxation process lead to significant deviations from the expectations and how do these compare to observations.
    \item Are the numerical models sufficiently stable over a wide range in cluster mass and time-scales.
\end{itemize}
} A good compilation of observed relations can be found in \citet{2011A&A...535A...4R}. Additionally we compare to recent results on the Compton-y parameter from \citet{2010A&A...517A..92A}.   \par
We set-up a sample of four galaxy clusters with masses 1.5, 1, 0.5, 0.1 $\times 10^{15} \ \mathrm{M}_\mathrm{sol}$, respectively. We sample the first cluster with $2 \times 10^6$ particles, the others with $2.5 \times 10^{5}$. These clusters are then evolved for 5 Gyr. Using our projection code Smac2 (Donnert et al. in prep) we obtain bolometric X-ray luminosity, Compton-y and spectroscopic temperature  maps from the box. This allows us to compare  the simulated temperature with the observed temperature in figure \ref{img_scaling} \citep{2004MNRAS.354...10M}. {  We do not take into account the emission from metal lines, which might be important at $T < 2.6 \,\mathrm{keV}$ \citep{1988xrec.book.....S}. } \par
The resulting scaling relations are shown in figure \ref{img_scaling}, from top left to bottom right: Bolometric X-ray luminosity vers. (spectroscopic) temperature, mass inside $r_{500}$ vers. temperature, mass inside $r_{500}$ vers. bolometric X-ray luminosity and Compton-Y parameter within $r_{500}$ vers. mass inside $r_{500}$. We mark the observed relations as a black line and the predicted correlation from equations \ref{eq.Tc}, \ref{eq.xray} and \ref{eq.sz} in brown. We overplot the evolution of the 4 clusters in red, blue, green and violet, respectively. In the $L_\mathrm{X}-T$ plot we add the relation predicted from the self-similar model as a black dotted line \citep{2011arXiv1112.5035B}. In the $L_\mathrm{x}-T$ relation we use the spectroscopic temperature from the simulation, which is a good approximation to the observed temperature. In the M-T relation we use the temperature from the simulation to make the relation self-consistent with the mass estimate. \par
Note that $M_{500}$ is not a direct observable, but requires modeling on the observer side. This introduces an additional bias into the correlations as we take $M_{500}$ directly from the simulation and do not apply this type of modelling. \par
We find an excellent fit of our analytic model and the simulated clusters to the $L_\mathrm{X}-T$ relation predicted by self-similar collapse. The observed correlation is steeper due to different morphological properties of small clusters, an effect we do not include in our model. Our model is still in the observed scatter for cluster masses above a few $10^{14}\,M_\mathrm{sol}$. \par
In the $M_{500} - T_{500}$ correlation the simulated clusters have slightly higher masses than predicted from analytics, probably from numerical effects during particle settling. Unlike in the $L_\mathrm{x}-T$ relation, we do not use the (lower) spectroscopic temperature estimate. The analytical model correlation and the simulated clusters agree very well with the observed sample. \par
In the $M_{500} - L_\mathrm{X}$ correlation the simulated clusters agree reasonably well with the sample. However the predicted model scaling is steeper than the observed scaling, which affects mostly small clusters. However the scatter of the observed clusters is quite large, so our model can still be considered reasonably well fitting. \par
The $Y_\mathrm{C,500} - M_{500}$ predicted by our model is very similar to the Y-M correlation derived from the universal pressure profile (dashed line) in \citet{2010A&A...517A..92A}. The simulated clusters are slightly offset, probably due to the numerical settling and fit the observed correlation and sample very well.\par
{ All numerical models show practically no evolution of observable quantities over 5 Myrs of time integration.}\par
We conclude that the scheme presented here models the observed correlations for galaxy clusters sufficiently well for most applications. Significant deviations occur only for clusters with $M_{200}$ smaller than a few $10^{14} \, M_\mathrm{sol}$, where real clusters are more strongly affected by merging. A good fit is obtained for the two important observable correlations $L_\mathrm{X}-T$ and $Y_\mathrm{C,500} - M_{500}$. The two other correlations are fit with an offset, due to numerical effects and differences in mass estimates between the model/simulation and the observations. \par

\subsection{Discussion}
Our model represents a \emph{compromise} between numerical practicality and realism. It is tuned to fit the two most important scaling relations $L_\mathrm{x}-T$ and $Y_{\mathrm{C}, 500}-M$. To summarise, we made the following assumptions: 
\begin{itemize}
    \item { The DM density follows a Hernquist profile with a slope of $-1$ in the centre of the cluster and a slope of $-4$ at the outskirts. This represents not only a viable fit to observations, but provides the ''naturally'' truncated mass profile need for this kind of simulations.}
	\item The outer slope of the gas density profile is $-1$. This makes the hydrostatic equation analytically treatable. However the resulting outer slope of the universal pressure profile turns out flatter than observed (fig. \ref{img_model}, bottom right). This leads to an inverted baryon fraction in the boundary region, which however is several virial radii away from the cluster centre.  
	\item The ICM core radius is a fixed fraction of the dark matter halo scale length. Therefore the model consistently fits the self-similar L-T relation. Significant deviations from the observed relations occur only for small cluster masses. This is also consistent with the pressure profile, where small clusters show larger deviations from the observed fit.
	\item The merger is embedded in a periodic box with a side-length dependent on $r_{200}$ of the main cluster. 
\end{itemize}
Deviations of the analytical model from the observed profile and scalings are caused mostly by the choice of the gas profile. We have refrained from introducing another mass dependence in baryon fraction and core radius, because it does not yield any substantial improvement and would only further complicate the model. Naturally we can not exclude the possibility that an even more convenient functional form for the gas density exists, which fits the observations even better. Nonetheless our model provides a useful, well-defined tool to investigate binary cluster mergers.

\section{Cluster Merger - 'El Gordo'}\label{merger}
The cluster ACT-CT J0102-4915, nickname 'El Gordo' is a large galaxy cluster at redshift $z = 0.870$ undergoing a merger \citep{2012ApJ...748....7M,2013ApJ...770L..15Z}. Its total X-ray luminosity in the 0.5 - 2 keV band is observed as $L_\mathrm{X} = 2.19 \pm 0.11 \times 10^{45} h_{70}^{-2} \,\mathrm{erg/s} $. A preliminary weak lensing analysis\footnote{See also the slides of J. Hughes talk held at the SnowCluster meeting in March 2013} (Jee et al. in prep) gives a distance of the DM cores of 700 kpc. Due to its high mass ($M_\mathrm{200} \approx 2.8 \times 10^{15}\,M_\mathrm{sol}$) and early formation it represents are rare event in the $\Lambda$CDM model and is therefore hard to simulate in the cosmological context. This makes it an ideal target for our numerical model. Furthermore its peculiar comet-like shape and the fluctuations in the X-ray brightness of the tail of the comet are not fully understood yet. \par

\subsection{Model}

\begin{table} 
    \centering
    \begin{tabular}{r | c c}
    Parameter &  Cluster 0  & Cluster 1  \\\hline
    $M_{200}$ & $1.9 \times 10^{15}\, M_\mathrm{sol}$ & $8.1 \times 10^{14}\, M_\mathrm{sol}$\\
    $c_\mathrm{NFW}$ & 2.9 & 3.2 \\
    $v_\mathrm{CoM}$ & 300 km/s & -2300 km/s \\
    $\beta$ &   2/3 &   2/3  \\
    $r_\mathrm{core}$ & 300 kpc & 25 kpc \\
    $r_{200}$ & 2550 kpc &  1925 kpc \\ 
    $d_\mathrm{CoM}$ & -1573 kpc & 3658 kpc \\
    \end{tabular}
    \caption{Model parameters for the simulated El Gordo cluster, after \citep{2012ApJ...748....7M}.}
    \label{tab.eg_params}
\end{table}
We model El Gordo following the parameters estimated in \citet{2012ApJ...748....7M} from observations. The initial conditions have a total mass in $r_{200}$ of $M_{200} = 2.7 \times 10^{15}\,\mathrm{M}_\mathrm{sol}$, a baryon fraction of 0.17 (in $R_{200}$) and a mass ratio of 0.43. The smaller of the two clusters is modelled as cool-core cluster, with only a small core of 25 kpc. We use a small impact parameter of 20 kpc to break perfect symmetry in the simulation. The clusters are set roughly on a zero energy orbit at a distance to match their radial density profiles. A summary of individual cluster parameters is given in table \ref{tab.eg_params}.  \par

\subsection{Implementation}
\begin{figure}
    \centering
    \includegraphics[width=0.4\textwidth]{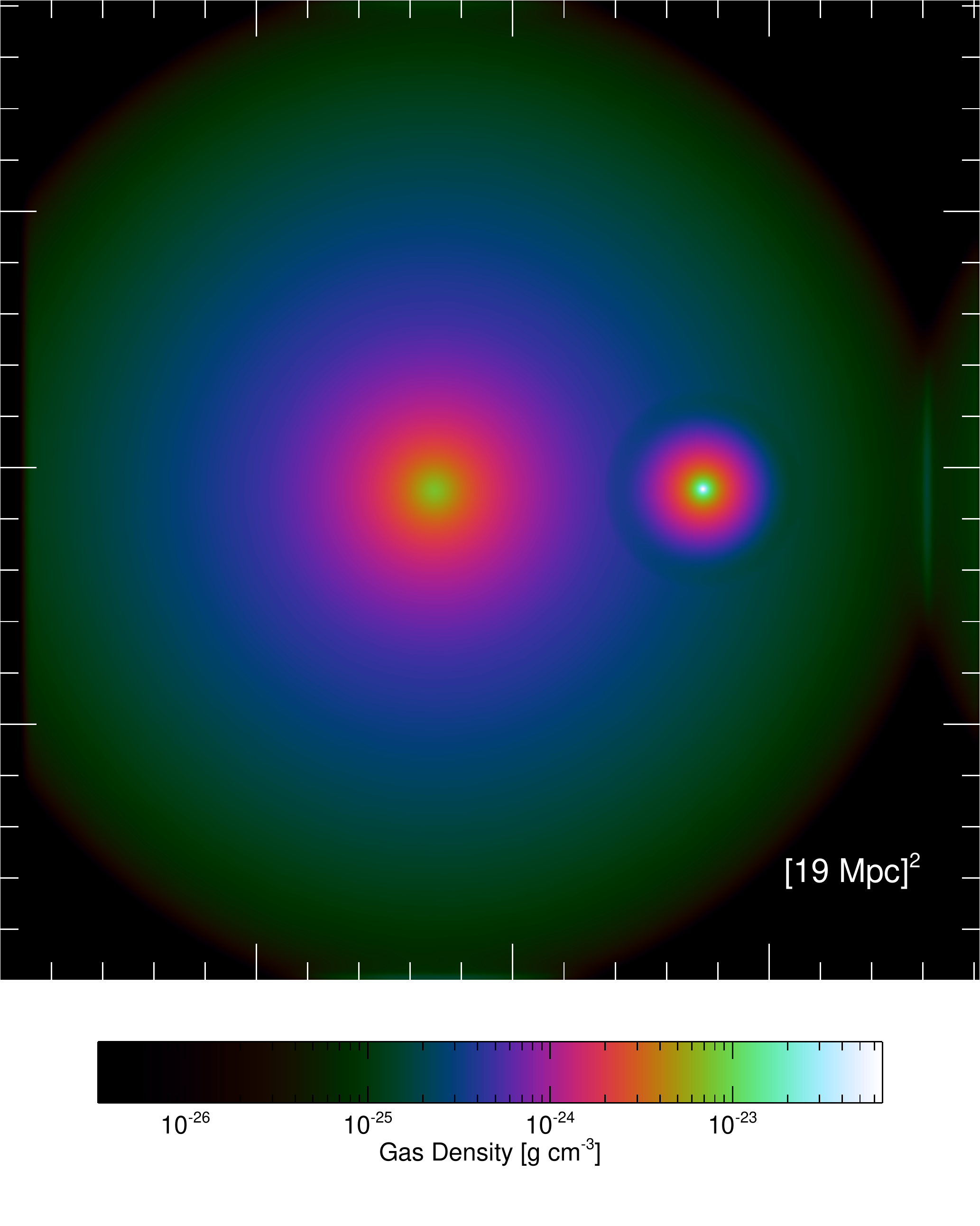}
    \caption{Projection of the gas density distribution of the initial conditions. Shown is the whole periodic box with a side length of 19122 kpc, containing 50 Million SPH particles.}\label{img.eg_ic}
\end{figure}

To sample the cluster we use 50 Million SPH and collisionless DM particles, respectively. This gives an effective mass resolution of $2.6\times10^{7}\,M_\mathrm{sol}$ for SPH and  $6.9\times10^{7}\,M_\mathrm{sol}$ for DM particles. Cluster 0 is sampled out to the boxsize of the simulation, while cluster 1 is sampled up to its $r_{200}$. \par
We relax both clusters separately in periodic boxes with high artificial viscosity and thermal conduction to get rid of sampling noise. Following our prescription in section \ref{sect.box} the merger is setup in a periodic box of size $L_\mathrm{Box} = 19122 \,\mathrm{kpc}$, centred on the center of mass of the system. We resample all particles of cluster 0 that overlap with cluster 1 outside of $r_{200}$ of cluster 0. \par
The two clusters are then merged in a comet like shape, i.e. all particles receive the velocity of the small cluster that fulfil one of the two conditions:
\begin{align}
   \sqrt{\left(\vec{r}_i - \vec{r}_1\right)^2 } \le r_{200,1}, \\
   \left( (y_i - y_1)^2 + (z_i - z_1)^2 \right)^{1/2} \le r_{200,1}
\end{align}
where $\vec{r}_i = (x_i, y_i, z_i)^T$ is the position of particle $i$, $\vec{r}_1 = (x_1, y_1, z_1)^T$ the position of the center of the small cluster and $r_{200,1}$ the virial radius of the small cluster. All other particles are set to the velocity of the large cluster.\par

\subsection{Results}
\begin{figure*}
    \includegraphics[width=\textwidth]{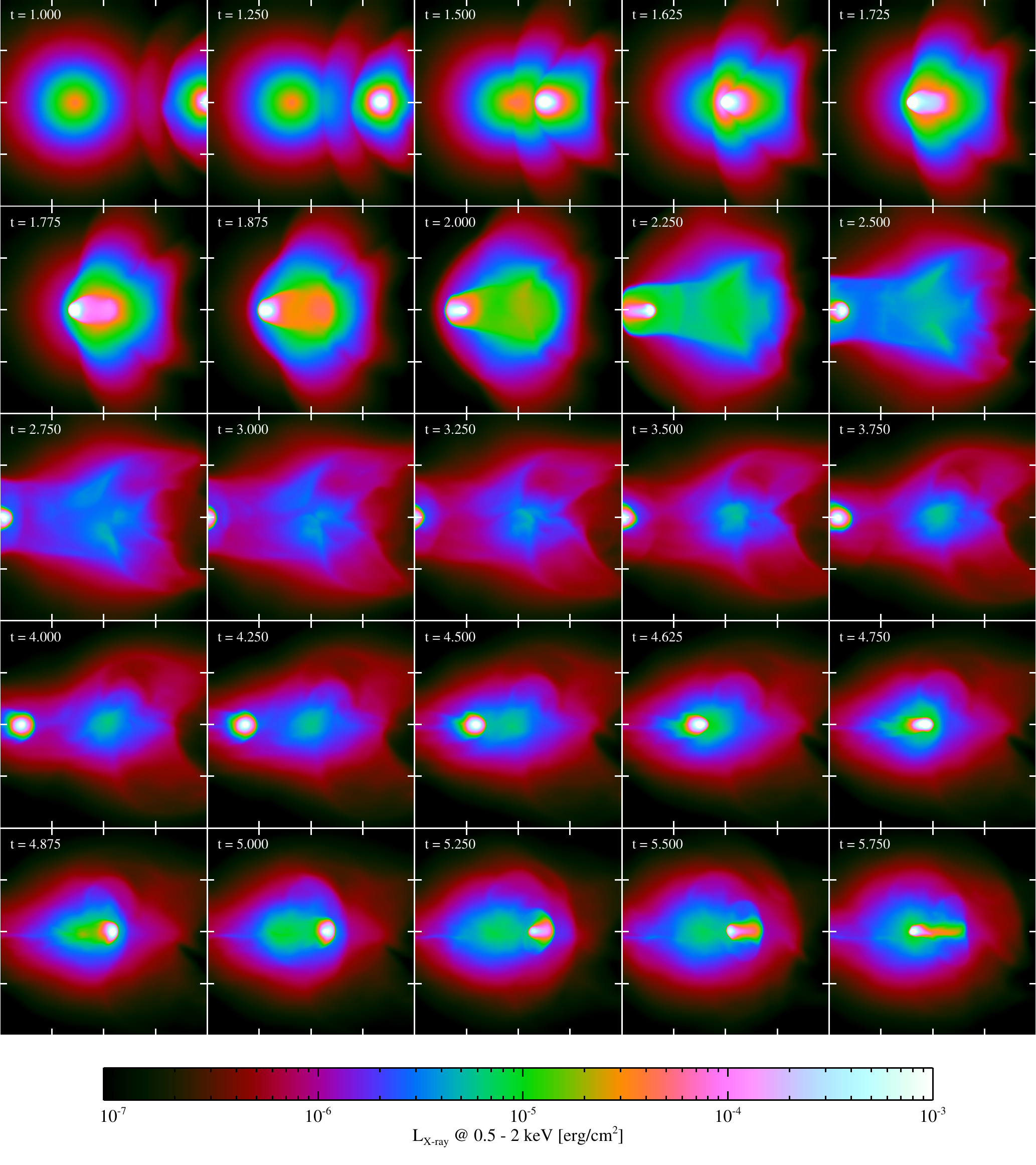}
    \caption{Projections of the X-ray brightness of the system at different stages of the merger. Each panel covers 2 x 2 Mpc. The image is centered on the center of mass of the system.}\label{img.eg_Lxevo}
\end{figure*}

The simulation was run on the Fermi BlueGene/Q machine at CINECA using 256 MPI tasks with 64 OpenMP threads each. A total of 200000 CPU hours were consumed for relaxation and simulation. To our knowledge this is the highest-resolution simulation of this kind carried out to date. We evolve the adiabatic run for 6 Gyrs, taking a snapshot every 25 Myrs.  We produce projections of X-ray luminosity, DM density and  Sunyaev-Zeldovich effect for every snapshot. A movie of the simulation is available from the author upon request via email. In figure \ref{img.eg_Lxevo}  we show a projection of the X-ray brightness for selected snapshots. \par
The system undergoes three main merging passages within the simulated 6 Gyrs. Upon infall a shock develops. Around $t=1.75$ Gyr, shortly after the first passage the system passes through the 'observed phase'. A Gyr after the initial passage the main cluster ICM shows large scale turbulent motions. After the second passage the small dark matter core is oscillating on increasingly shorter time scales within the gravitational potential of the larger cluster, stirring turbulent motions. 

\begin{figure}
    \includegraphics[width=0.5\textwidth]{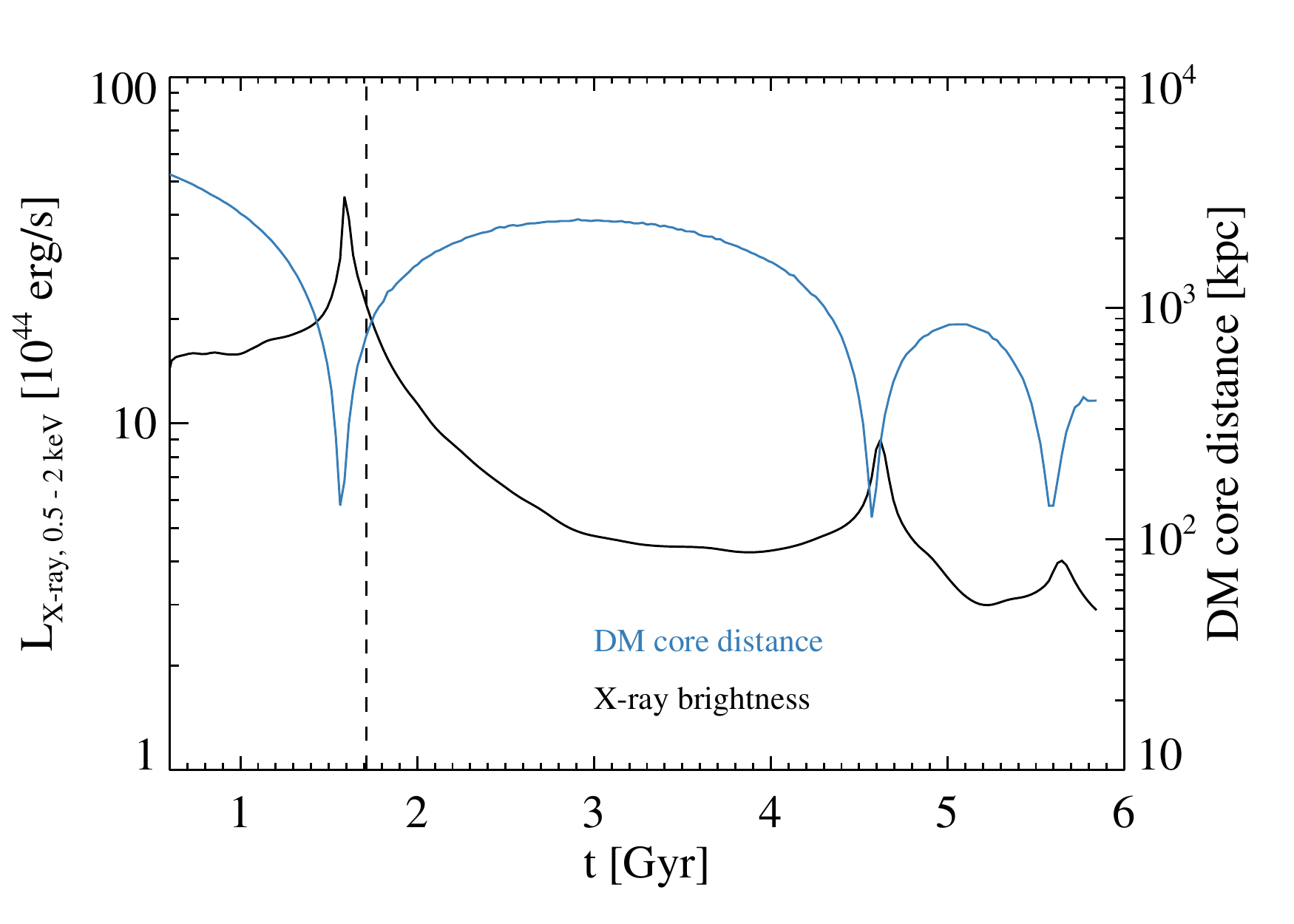}
    \caption{Time evolution of the X-ray luminosity of the system in the 0.5 to 2.0 keV band (black). Distance of the two dark matter cores (blue). Note that the DM core distance does not reach 0 because of the finite time resolution (number of snapshot) used to make the graph.}\label{img.eg_Lx_t}
\end{figure}
In figure \ref{img.eg_Lx_t} we show the evolution of the X-ray luminosity in the ACIS band of 0.5 to 2.0 keV. We also add the distance of the two dark matter cores in blue. The system undergoes three mergers within 6 Gyr, with X-ray brightness reaching local maxima and DM core distance approaching zero. The total X-ray brightness systematically declines, probably because the merger adds heat and bulk motions to the ICM. These additional sources of pressure 'puff up' the cluster atmosphere, which drastically reduces the brightness, because of its strong dependency on density. The position of the DM core distance minimum and the X-ray maximum suggests that the displacement between DM cores and X-ray peak is only small in this cluster.

\subsubsection{Comparison to the Observed Cluster}
\begin{figure*}
    \includegraphics[width=0.49\textwidth]{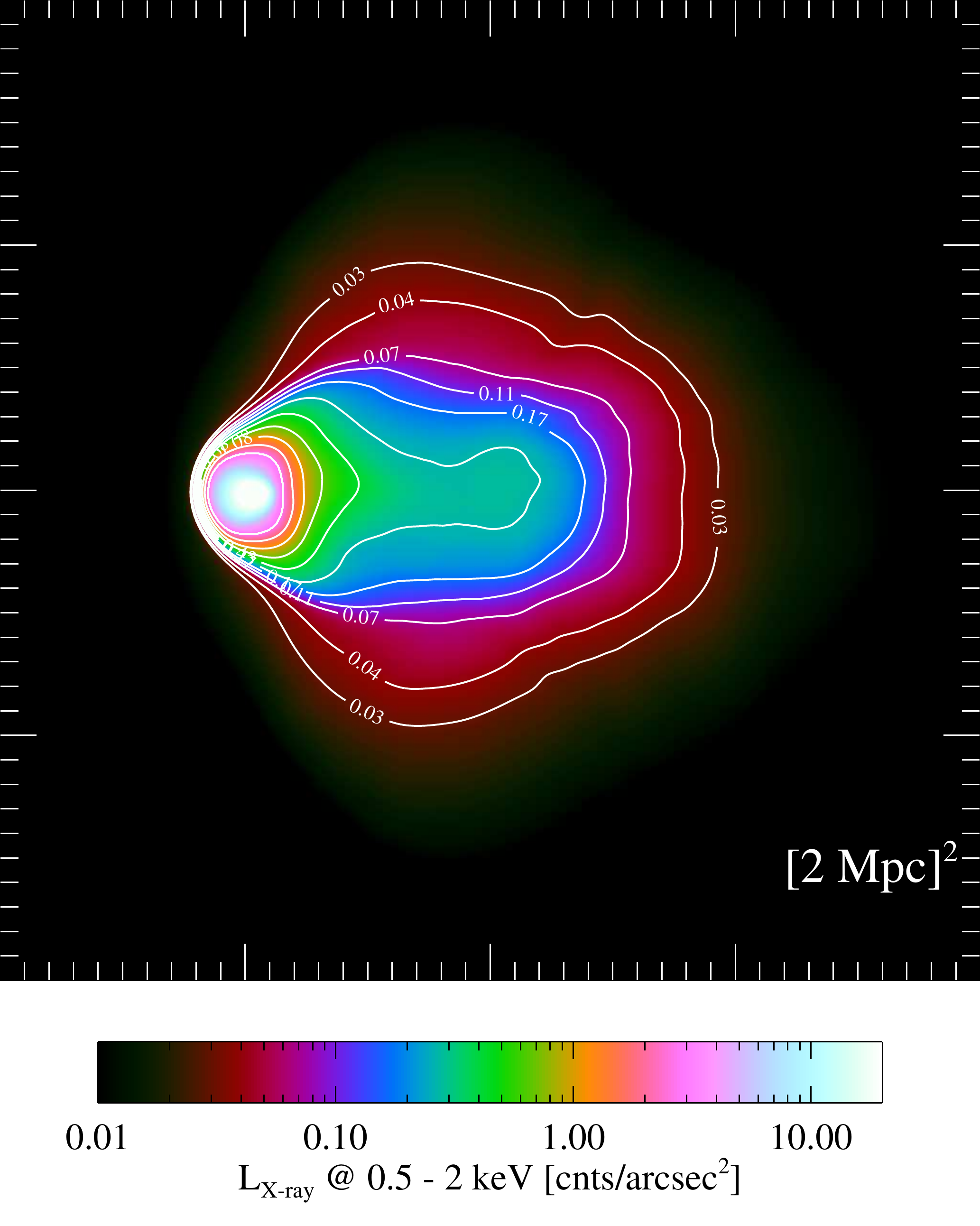}
    \includegraphics[width=0.49\textwidth]{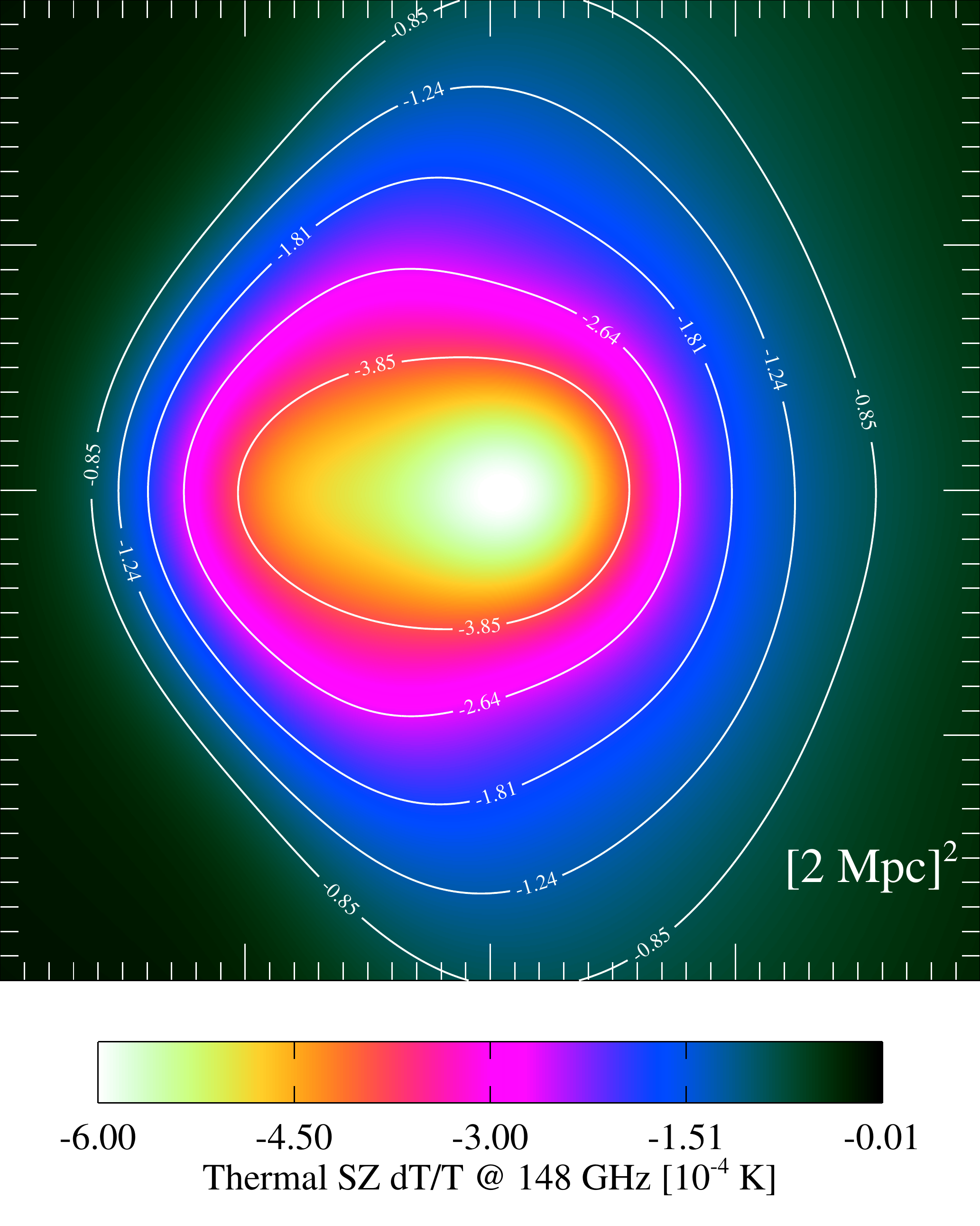} \\
    \includegraphics[width=0.49\textwidth]{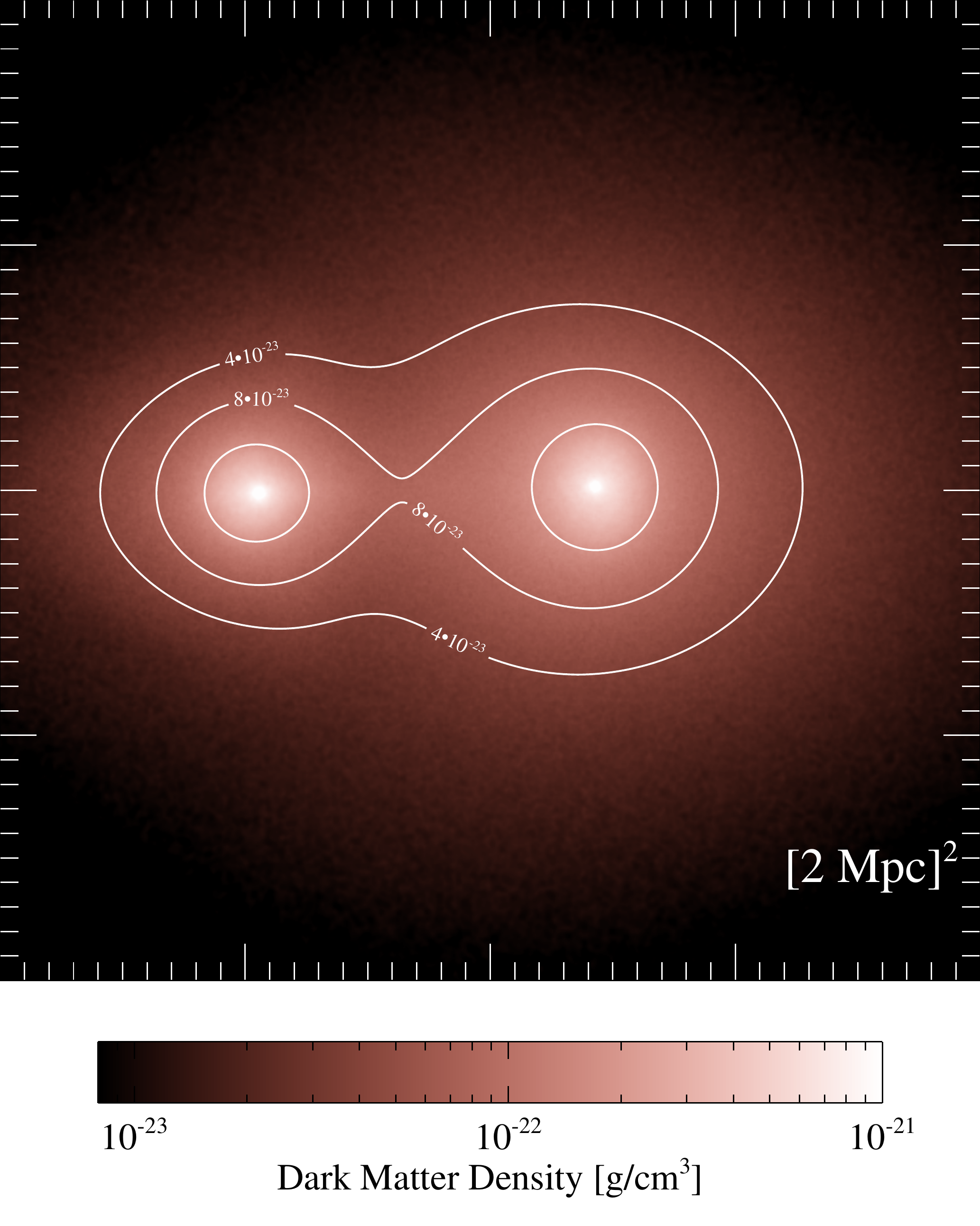}
    \caption{Projections of Xray brightness in the 0.5-2.0 keV band, thermal SZ effect, DM density (top left to bottom) with two Mpc side length. The X-ray image was converted to ACIS counts using $5.8 \times10^{12} \,\mathrm{erg}/\mathrm{cnt}$. The contours are logarithmically spaced between $0.03$ and $2.71\,\mathrm{counts}/\mathrm{arcsec}^2$. The SZ image at 148 GHz is smoothed with a Gaussian corresponding to 1.4 arcmin at z=0.87. The contours are logarithmically spaced between $-385\,\mu\mathrm{K}$ and $-85\,\mu\mathrm{K}$. The dark matter density projection features three contours at $4$, $8$, $20 \times 10^{-23} \,\mathrm{g}/\mathrm{cm}^3.$ 
 }\label{img.eg_portrait}
\end{figure*}
The best point in time to compare the simulation with the observed cluster can be identified from the distance of the DM cores and the total X-ray luminosity of the system (dashed line in figure \ref{img.eg_Lx_t}). We do not consider the morphology of the X-ray emission at this point, because it may include substructure, which we are not able to account for in our model. For an observed DM core distance of 700 kpc and an observed X-ray luminosity of $L_\mathrm{X} = 2.19 \times 10^{45} \,\mathrm{erg/s}$ we find that the snapshot 70, $t = 1.75$ Gyr after simulation start compares best. Here the DM cores are 690 kpc apart and X-ray luminosity is $L_\mathrm{X} = 2.2 \times 10^{45} \,\mathrm{erg/s}$. Considering the differences in morphology the X-ray luminosity has to be taken with a grain of salt. The modeling uncertainties are certainly in range of $5 \times 10^{44} \,\mathrm{erg/s}$, because $L_\mathrm{X} \propto n_\mathrm{th}^2$ and most of the emission is concentrated in the cool core. The DM core distance however can be considered a more solid prediction, given the mass estimates are roughly correct. Note that access to the observed X-ray maps would allow more precise modelling of the cool core cluster, yielding a better model. \par
In figure \ref{img.eg_portrait} we show projections of X-ray flux, SZ-effect, DM density (top left to bottom). We converted the X-ray brightness between 0.5 and 2 keV to ACIS counts\footnote{via WEBPIMMS} using $5.8 \times 10^{-12}\, \mathrm{erg}/\mathrm{cnt}$. Here logarithmic contours are between $0.03$ and $2.71\,\mathrm{counts}\, \mathrm{arcsec}^{-2}$. The SZ effect was computed at 148 GHz and overlaid with five logarithmically-spaced contours between $-385\,\mu\mathrm{K}$ and $-85\,\mu\mathrm{K}$. We add three contours to the DM density projection, at $4$, $8$ and $20 \times 10^{-23} \,\mathrm{g}/\mathrm{cm}^3.$ \par
In the X-rays the overall morphology of the simulated cluster agrees well with the observed system. However the emission is more concentrated on the cool core in the simulation compared to observations \citep[compare to ][ figure 1 bottom left]{2012ApJ...748....7M}. The simulated system shows a smooth flow in the wake at this point in time. The cool core has lost less mass.   \par
The smoothed SZ signal of the simulation is dominated by the bulk of the gas
which is bound to the DM halo of the parent cluster. The signal is  consistent
with their  SZ signal from the deprojected X-ray data \citep{2012ApJ...748....7M}. The direct observation is rounder, probably due to the low resolution of the instrument. \par
The simulation, as the observations, show a significant offset between the peak of the SZ signal and the peak of the X-ray signal. This is not uncommon for a cluster of this size \citep{2001PASJ...53...57K} and can be used to measure high temperatures in the shock region of the cluster \citep{2004PASJ...56...17K}.\par
At the point in time chosen here ($t = 1.75\,\mathrm{Gyr}$), the passage of the cool core drags along ICM plasma from the main cluster, because both cluster atmospheres are just about to separate from each other. \par
Our model represents an idealised case, i.e. absence of substructure and injection of turbulence only at shear flows and through displacement and collapse.  
\begin{figure*}
    \includegraphics[width=0.75\textwidth]{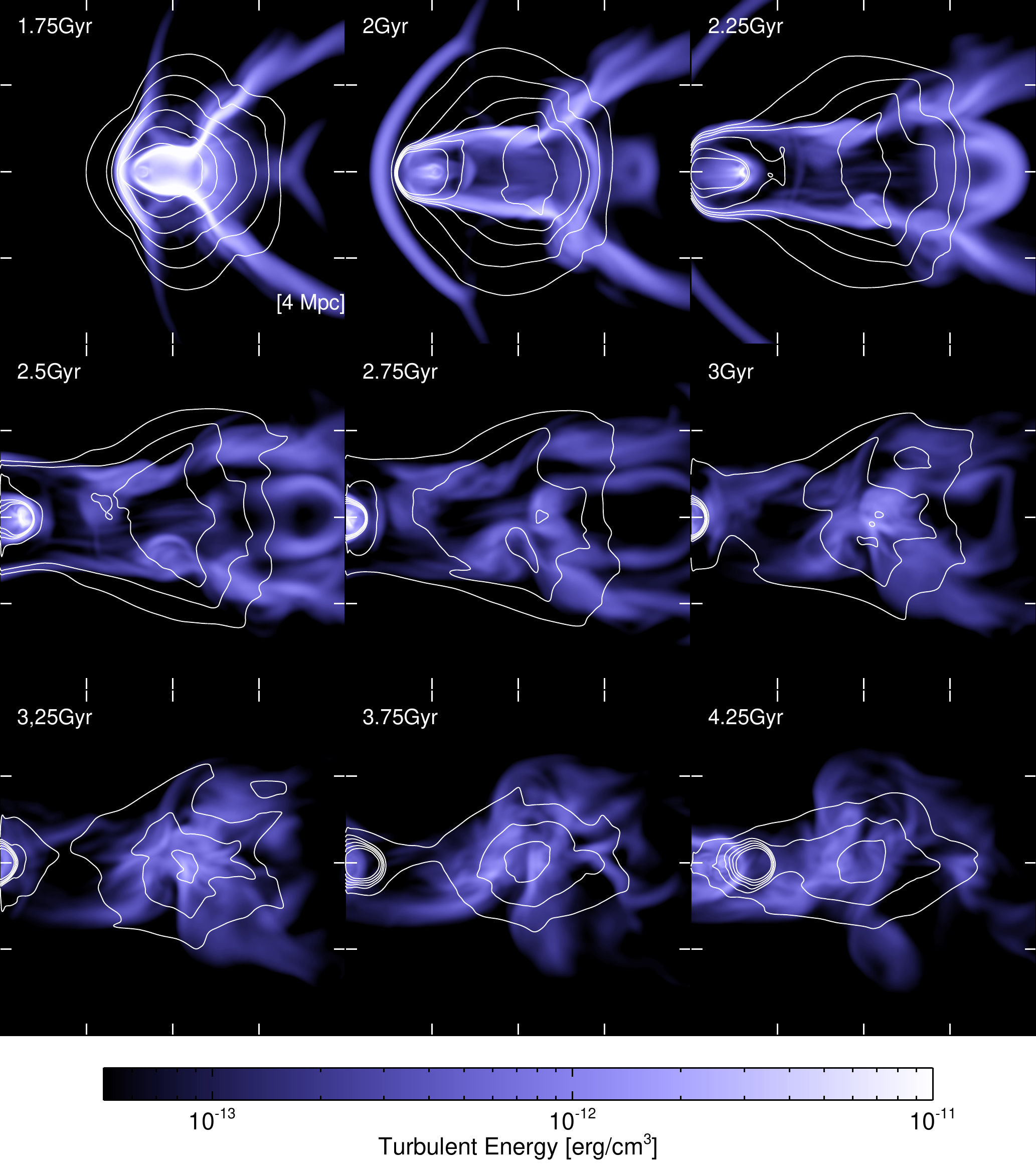}
    \caption{Projections of the turbulent energy from the kernel weighted velocity dispersion at 9 times between 1.75 Gyr and 4.25 Gyr. The size of each panel is $(4 \,\mathrm{Mpc})^2$. We overplot X-ray contours in arbitrary units.}\label{img.eg_turb}
\end{figure*}
{ Therefore the separation of the two cluster atmospheres results in a smooth flow. Cluster-wide turbulence sets in later in our simulation, roughly 1 Gyr
after the first core passage, at $t = 2.75$ Gyr. To illustrate this point we consider the turbulent energy $E_\mathrm{turb} = \rho \sigma^2$, with $\sigma$ the kernel weighted velocity dispersion. We first compute the kernel weighted velocity for every particle in the simulation. This effectively suppresses possible SPH noise from the estimate. Then we compute the RMS of these velocities from all neighbours of an SPH particle to obtain $\sigma$. This quantity measure compressive as well as solenoidal velocity components. \par
Figure \ref{img.eg_turb} shows a projection of the turbulent energy of a 500 kpc thick slice through the inner 4 Mpc of the simulation box. We add X-ray contours in arbitrary units to visualise the density structure.  At the 'el Gordo' stage (top left) the cluster is dominated by shocks and compression in the inner Mpc of the simulation. Half a Gyr later (top right), turbulent motions have not set in yet. Only 1 Gyr after the 'el Gordo' stage (centre) the ICM of the main halo is collapsing into the DM potential and complex velocity structures develop. These are still present at the second core passage (bottom right). }
This delayed onset of large-scale turbulent motions has been seen in early grid-based simulations of similar kind as well \citep{1999ApJ...518..594R}.\par

\subsection{Resolving Turbulence in the Wake}
{There have been recent claims that SPH may be unable to resolve turbulence due to kernel noise, subsequent gradient errors and the shape of kernel-based viscosity in idealised simulations of subsonic driven turbulence \citep{2012MNRAS.423.2558B}. This could in principle limit our ability to catch the bulk of the turbulence injected by the merger.\par
However \citet{2012MNRAS.420L..33P} have shown that subsonic turbulence in SPH is indeed \emph{converged} outside of the kernel aliasing scale\footnote{This scale is connected \emph{not} to the compact support / smoothing length of the kernel, but its smoothing scale, which is a kernel independent measure of of spatial resolution \citep[e.g.][]{2012MNRAS.425.1068D}}, even for the cubic kernel. We can therefore make the following estimate: In the wake of the cluster our simulations shows a WC6 compact support / smoothing length of $< 40$ kpc, which corresponds to a (kernel independent) smoothing scale / FWHM of $<8.16$ kpc. This corresponds to a compact support of $<30$ kpc for the cubic kernel. For this kernel the aliasing scale is roughly twice this value, which gives the \emph{most conservative} estimate of $60$ kpc, above which turbulent motions are resolved in the wake of our simulation. \par
Additionally we are using a different SPH kernel from \citet{2012MNRAS.423.2558B}, who used the cubic spline kernel with 64 neighbours. The WC6 kernel with 295 neighbours has larger compact support, i.e. takes into account more neighbours, while retaining the \emph{same} effective smoothing scale. This results in higher interpolation accuracy, similar to a higher order method for grid codes and improved gradient estimates (Beck et al. in prep.). Our choice of numerical viscosity parameters successfully surpresses noise in shocks, but reduces artificial viscosity away from shocks \citep{2005MNRAS.364..753D}. Here the obvious difference to the \citet{2012MNRAS.423.2558B} simulations can be obtained by a simple visual comparison of density projections \citep[see also][ fig 4.]{2012ASPC..453..249P}. \par
Furthermore our result of a \emph{delayed onset} of large scale  turbulence ($> 100$ kpc) is consistent with results from gridcodes, which report the onset of turbulence 1.3 Gyr after the first core passage \citep{1999ApJ...518..594R}. \par
We therefore conclude that our numerical method is able to resolve turbulent motions in the wake region down to scales of at least 60 kpc, well below the largest observed wake fluctuation on scales of several 100 kpc. 

\subsection{Discussion}

There are physical arguments that { motivate} the delayed onset of turbulence in our simulation. { However we stress that this is due to the idealised initial conditions and \emph{not} true for a full cosmological setup, where substructure and bulk motions are always present to some degree. We therefore do not see tension with these simulations.} \par

In the absence of substructure and pre-exisiting turbulent motions, the injection of turbulence is limited to two cases:
\begin{enumerate}
    \item The shear flow / KH instability can inject turbulence on small scales at the interface of the two cluster atmospheres. The injection scale will be of the order of the thickness of the shearing layer, much smaller than the observed fluctuations and limited to its edges.
    \item The bulk motion of the infalling cluster can inject turbulence on scales of the cluster core, which \emph{is} of the order of a few 100 kpc. However the pertubation / driving will act on a time scale of the core passage, $\approx 1 \,\mathrm{Gyr}$, due to the displacement and subsequent relaxation of the main clusters ICM. The core passage is not complete before 2 Gyrs after the beginning of the simulation. This is also expected from a back of the envelope estimate: For a relative infall velocity of 1500 km/s and a size of 2 Mpc, the injection timescale of turbulence is $\approx 1.3\,\mathrm{Gyr}$.  
\end{enumerate}
}
The absense of turbulence in the wake suggests that the characteristic appearance of the observed el Gordo cluster requires significant substructure and pre-existing turbulence in the parent cluster. This includes velocity, density and temperature fluctuations on scales of and below 100 kpc. Preexisting turbulence may be amplified by the merger. Substructure aids ablating and mixing the gas from the cool core through instabilities during in-fall. \par
In this view the parent cluster was not a relaxed system before the main merger event we observe today, but a highly disturbed system (closer to the 'puffed-up' simulated cluster 2 Gyr after the passage). \par
This is supported\footnote{Assuming that the X-ray background does not contribute significantly to the X-ray emission above $0.03 \,\mathrm{cnts}/\mathrm{s}$} also by the large extend and round shape of the observed system on scales of $> 1\,\mathrm{Mpc}$. Our simulated cluster is significantly smaller and appears less round in the X-rays. There also is a steep gradient in X-ray brightness in front of the outgoing core, which can not be flattened in projection by a rotation of the system on the plane of the sky. In contrast, the observed cluster shows significant emission ahead of the core. If we were to reproduce this shape and extend, the relaxed parent cluster ICM would have to be modelled with $r_\mathrm{c} > 400 \,\mathrm{kpc}$ which is unreasonably large even for such a massive system. \par
{ Given the size and early formation redshift one can argue that the parent cluster is likely to be disturbed. Objects of the size of the main el Gordo progenitor need of the order of 2 Gyr to relax. This would imply a formation redshift of $z > 1.4$ for a halo with mass of $>1.5 \times 10^{15}\, \mathrm{M}_\odot$, which is not expected in $\Lambda$CDM cosmology. Hence the main progenitor did not have enough time to relax, but underwent strong merging activity before the infall of the smaller halo. }
In this case the dynamics of the ICM flow during the merger will be highly irregular, inducing fluctuations which result in the characteristic wake. \par
Certainly a more sophisticated numerical model can be build. Here more detailed constrains from observations would help to improve the model as well as direct access to the X-ray and SZ maps for comparison. However this as well as non-thermal radio emission from the cluster is beyond the scope of this paper, but presents promising work for the future.

\section{Conclusions}\label{conclusions}

We have presented a model for isolated galaxy cluster mergers with an application to the 'El Gordo' cluster. We use the Hernquist profile to model the DM halo of the cluster. The ICM density distribution is assumed to be a $\beta$-model. We derive the temperature distribution for the hydrostatic equilibrium distribution in the case of $\beta = 2/3$. We gave approximate expressions for X-ray luminosity and the Compton-y parameter of such a cluster. A comparison of the model with the observed universal pressure profile and four observed scaling relations for galaxy clusters yields good agreement with the self-similar model. \par
We gave a detailed description of the methodology to implement this model in SPH simulations. We argued to simulate such cluster mergers in a periodic box to avoid numerical problems with escaping particles. We compared a number of numerical models of isolated clusters with observed scaling relations and found satisfactory agreement for this class of models. \par
We then applied our model to simulate the cluster ACT-CT J0102-4915 'El Gordo'. We showed how to setup such a system in SPH. Using the highest resolution simulation of the kind done so far we were able to reproduce the total X-ray luminosity and overall shape of the observed cluster reasonably well. We were not able to reproduce the characteristic fluctuations in the wake of the cluster using our smooth model. We argued that substructure is the cause of these fluctuations and that the parent cluster of 'El Gordo' was a highly disturbed system even before the major merger observed today. 

\section{Acknowledgements}\label{ack}

The author would like to thank the anonymous referee for useful comments that helped to sharpen the argumentation and improve the paper. Further thanks go to G. Brunetti, K. Dolag, A.Beck, R. Kale, and F. Vazza for comments on the paper. We thank K.Dolag and V.Springel for granting access to the MHD version of {\small GADGET}-3.
J.D. acknowledges support by the FP7 programme ’People’ of the European Union. Computations were performed at CINECA, with resources awarded through the ISCRA program to the project 'Simulations of MHD Turbulence in Mergers of Isolated Galaxy Clusters'.
\bibliographystyle{mn2e} \bibliography{master}

\label{lastpage}
\end{document}